\documentclass{emulateapj}

\usepackage{natbib}
\usepackage{ulem}
\usepackage{epsfig}
\usepackage{amsmath}

\shorttitle{Cooling, Feedback and Star Formation in A1664}
\shortauthors{Kirkpatrick et al.}
\slugcomment{Accepted for publications in The Astrophysical Journal}

\begin{document}

\title{A Chandra X-ray Analysis of Abell 1664: Cooling, Feedback and Star Formation in the Central Cluster Galaxy}

\author{C.~C. Kirkpatrick\altaffilmark{1}, 
              B.~R. McNamara\altaffilmark{1,2,3},  
              D.~A. Rafferty\altaffilmark{4}, 
              P.~E.~J. Nulsen\altaffilmark{3},
              L. B\^{i}rzan\altaffilmark{4},
              F. Kazemzadeh\altaffilmark{1},
              M.~W. Wise\altaffilmark{5},  
              M. Gitti\altaffilmark{6}
              and K.~W. Cavagnolo\altaffilmark{1}}






\altaffiltext{1}{Department of Physics \& Astronomy, University of Waterloo, 200 University Ave. W., Waterloo, ON N2L 3G1, Canada}
\altaffiltext{2}{Perimeter Institute for Theoretical Physics, 31 Caroline St. N., Waterloo, ON N2L 2Y5, Canada}
\altaffiltext{3}{Harvard-Smithsonian Center for Astrophysics, 60 Garden St., Cambridge, MA 02138}
\altaffiltext{4}{Department of Astronomy \& Astrophysics, The Pennsylvania State University, 525 Davey Laboratory, University Park, PA 16802}
\altaffiltext{5}{Astronomical Institute ``Anton Pannekoek", University of Amsterdam, Krusilaan 403, 1098 SJ Amsterdam, The Netherlands}
\altaffiltext{6}{INAF - Osservatorio Astronomico di Bologna, via Ranzani 1, I-40127 Bologna, Italy}

\begin{abstract}
The brightest cluster galaxy (BCG) in the Abell 1664 cluster is unusually blue and is forming stars at a rate of $\sim$ 23 $M_{\sun}$ yr$^{-1}$.  The BCG is located within 5 kpc of the X-ray peak, where the cooling time of 3.5$\times$10$^8$ yr and entropy of 10.4 keV cm$^2$ are consistent with other star-forming BCGs in cooling flow clusters.  The center of A1664 has an elongated, ``bar-like" X-ray structure whose mass is comparable to the mass of molecular hydrogen, $\sim$ 10$^{10}$ $M_{\sun}$ in the BCG.  We show that this gas is unlikely to have been stripped from interloping galaxies.  The cooling rate in this region is roughly consistent with the star formation rate, suggesting that the hot gas is condensing onto the BCG.  We use the scaling relations of \citet{bir08} to show that the AGN is underpowered compared to the central X-ray cooling luminosity by roughly a factor of three.  We suggest that A1664 is experiencing rapid cooling and star formation during a low-state of an AGN feedback cycle that regulates the rates of cooling and star formation.  Modeling the emission as a single temperature plasma, we find that the metallicity peaks 100 kpc from the X-ray center, resulting in a central metallicity dip.  However, a multi-temperature cooling flow model improves the fit to the X-ray emission and is able to recover the expected, centrally-peaked metallicity profile.
\end{abstract}

\keywords{cooling flows --- galaxies: clusters: individual: A1664 --- galaxies: starburst --- X-rays: galaxies: clusters}

\section{Introduction}

The fate of hundreds to thousands of solar masses per year of gas thought to be condensing out of the X-ray atmospheres of galaxy clusters is a problem that has puzzled astronomers for more than three decades \citep{fab94}.  New {\it Chandra} and XMM-{\it Newton} observations have recast the problem in terms that have significant consequences for understanding the formation of galaxies and super-massive black holes.  First, high resolution spectra of the thermal emission from cooling flows failed to show the characteristic recombination lines from highly ionized metals in the condensing gas \citep[for a review]{pet03,pet06}.  The XMM-{\it Newton} observations did not rule out cooling altogether \citep[eg.,][]{san08}, but they were the first to show convincingly that the condensation rates must lie far below the levels predicted by pure cooling models.  Second, images of cluster atmospheres revealed large-scale cavities and shock fronts associated with powerful AGN outbursts in central dominant galaxies \citep[see][for a review]{bmc07}.  Measurements of cavity sizes and their surrounding pressures provide a convenient and reliable gauge of  the mechanical $pV$ work (energy) expended by radio jets as they inflated cavities against the surrounding gas pressure \citep{bmc00}.  Assuming the cavities are driven to their current locations primarily by buoyant forces \citep[eg.,][]{chu02}, the mean jet power is comparable on average to the power required to quench cooling in the cores of galaxies and clusters \citep{bir04,dun06,raf06}.  Apparently, the supermassive black holes located in the nuclei of BCGs combined with abundant fuel accreting onto them provide a natural and powerful feedback mechanism that is energetically able to maintain most (but not all) of the cooling gas above a few keV \citep{bru02,rey02,dal04,rus04a,rus04b,bru05,hei06,ver06}.  Thus the failure of XMM-{\it Newton} to find strong cooling lines can be attributed largely to AGN feedback.  How this jet power heats the gas and the extent to which other heating mechanisms such as thermal conduction, galaxy mergers, etc. are aiding it is poorly understood, and remains an outstanding issue \citep{zak03,dol04,voi04,den05,poo08}.

Evidence for residual cooling can be inferred by the unusually high star formation rates and reservoirs of cold gas found in BCGs.  Although this star formation is frequently attributed to stripping from the occasional gas-rich galaxy \citep[eg.,][]{hol96}, several surveys undertaken over the last few decades have tied the presence of nebular line emission, molecular gas, and star formation to cooling flows \citep{john87,bmc89,cra95,cra99,card98,edge02,sal03,raf06,don07,love07,bil08,ode08,raf08}.  Despite these ties, the enormous gap between the cooling and star formation rates raised very serious issues.  This situation again changed dramatically when XMM-{\it Newton}'s downward-revised condensation rates were shown to be in near agreement with the star formation rates in many, but not all, systems \citep{bmc02,hic05,raf06,ode08}.

Recently, evidence tying the cooling of hot halos to star formation was found by \citet{raf08}, who showed that star formation ensues when the central cooling time falls below a remarkably sharp threshold value of $\sim$ 5$\times$10$^8$ yr or equivalently a central entropy of 30 keV cm$^2$.  This and two other criteria:  the X-ray and galaxy centroids lie within $\sim$ 20 kpc of each other, and the X-ray cooling luminosity exceeds the jet (cavity) power, appear to govern the onset of star formation in cooling flows.  \citet{cav08} found a similar threshold for the onset of H$\alpha$ and radio emission \citep[see also][]{hu85}.  The H$\alpha$ threshold is certainly related to but not identical to the star formation threshold, as we show in this paper.  These new results are difficult to understand in the context of mergers or stripping, but instead they tie the presence of star formation and AGN activity closely to cooling instabilities in the hot atmospheres \citep{raf08,voit08,sok08}.

These developments have broad implications for understanding the formation and evolution of galaxies.   Outflows driven by gas accretion onto massive black holes may have regulated the growth of bulges giving rise to the observed correlation between bulge mass and super-massive black hole mass in nearby galaxies \citep{fer00,geb00,har04}.  In simulations, AGN feedback at late times in the so-called ``radio mode," of which feedback in cooling flows is thought to be the archetype, is able to suppress cooling and star formation in the hot halos of giant elliptical galaxies and BCGs \citep{bow06,cro06}.  This process may explain the turnover at the bright end of the galaxy luminosity function and the lack of bright blue galaxies expected in standard $\Lambda$CDM models.  Cooling flows permit study of this process in detail.

\object{Abell 1664} is a good candidate for studying star formation and cooling in galaxy cluster cores.  It is a cooling flow cluster \citep{all95} located at redshift $z=0.128$ \citep{all92} and has an Abell richness class of 2 \citep{abe89}.  Based on ROSAT data, the X-ray luminosity in the 0.1-2.4 keV band is 4.1$\times$10$^{44}$ erg s$^{-1}$ \citep{all92}.  A1664 hosts a BCG known to be a very bright H$\alpha$ emitter with a luminosity of at least 1.6$\times$10$^{42}$ erg s$^{-1}$, and forming stars at a rate of 14-23 $M_{\sun}$ yr$^{-1}$ \citep{wil06,ode08}.  In the X-ray waveband, the cluster is regular in shape on large scales.  The core shows signs of a surface brightness break along the eastern edge at a radius of 30$''$ from the center and an elongated surface brightness peak at the center.

In this paper, we present our analysis of the recent {\it Chandra} observations and deep $U$- and $R$-band images of the cluster and its core.  Section 2 describes the data set and reduction process.  In Section 3, we describe our analysis of the X-ray and optical data.  In Section 4 we interpret the data.  Finally, we summarize in Section 5.  Throughout the paper we assume a $\Lambda$CDM cosmology with H$_{0} =$ 70 km s$^{-1}$ Mpc$^{-1}$, $\Omega_{M} =$ 0.3, and $\Omega_{\Lambda} =$ 0.7.  The angular scale is 2.29 kpc arcsec$^{-1}$ at the redshift of the cluster.  All uncertainties quoted are 90\% confidence intervals.

\section{Observations \& Data Reduction}

Abell 1664 was observed for 37 ks on 2006 December 4 (ObsID 7901) using the ACIS-S3 CCD on-board the {\it Chandra X-ray Observatory}.  The focal plane temperature during the observation was kept at -120\degr C.  CIAO version 3.4 was used for all data reduction and preparation with version 3.4 of the calibration database.  Standard filtering was applied to eliminate background flares, producing a loss of 0.4 ks from the total exposure time.  Point sources have been identified and removed using {\it wavedetect} and confirmed by eye.  Time-dependent gain and charge transfer inefficiency corrections have been applied.  Blank-sky background files used for background subtraction were normalized to the source image count rate in the $9.5-12$ keV band.  When extracting spectra from a region, a background spectrum is extracted from the equivalent region of the blank-sky background event file.

Optical images of the cluster's BCG were obtained with the Kitt Peak National Observatory's 4 m telescope on 1995 February 2 using the T2KB CCD camera.  The seeing at the time was approximately 1.3$''$.  Images were taken through $U$ and Gunn $R$ filters with total exposure times of 2400 s and 800 s, respectively.  These filters were chosen to avoid contamination by emission lines from [OII] 3272 and H$\alpha$ 6563 at the redshift of the cluster.  Additional imaging was done using an $R$-band filter that includes H$\alpha$ emission with a total exposure time of 800 s.  The individual images have the bias level subtracted and have been flat-fielded using twilight sky exposures.  Finally, the images for each band were combined into single $U$ and $R$ images and color corrected for use in our analysis.  An H$\alpha+$[NII] emission map was created by taking the scaled difference between the normalized $R$- and Gunn $R$-band images.

\section{Data Analysis}

\subsection{Images}

We present in Figure 1 the $R$-band DSS image of A1664.  Overlaid are the X-ray contours spanning the entire ACIS-S3 CCD (8$'$ $\times$ 8$'$).  On large scales, the X-ray emission is fairly regular, but with asymmetries to the south and east.  The inner region is more complex.  The cluster core image in the 0.4 - 7.0 keV band is presented in the upper left panel of Figure 2.  A bar-like structure is shown as the dark region approximately 22.5$''$ $\times$ 10$''$ (52 $\times$ 23 kpc) in size associated with the BCG, marked A.  To the east, a brightness edge, marked B, is seen at a radius of 30$''$ from the cluster center.  In the upper right panel of Figure 2, H$\alpha$ emission is aligned with the ``bar" region, though the bar is more extended than both the H$\alpha$ and star forming region.  The feature marked C is thought to be a disturbance created by a neighboring galaxy \citep{wil06}.  In the bottom left and right panels of Figure 2 we present the $R$- and $U$-band images, respectively.  The $R$-band shows that the BCG's core does not share the same position angle on the sky as the outer envelope.  In the $U$-band, there is more structure.  There are two distinct clumps north and south of the center.  We will discuss possible origins of these features in the following sections.

\subsection{X-ray Properties}

In the following sub-sections, we present our analysis of the {\it Chandra} observation of A1664.  Our purpose is to obtain the mean properties of the cluster including the gas temperature, density, metallicity, cooling rate, and how these properties vary with radius.

\begin{figure}[t]
\epsscale{1.15}
\plotone{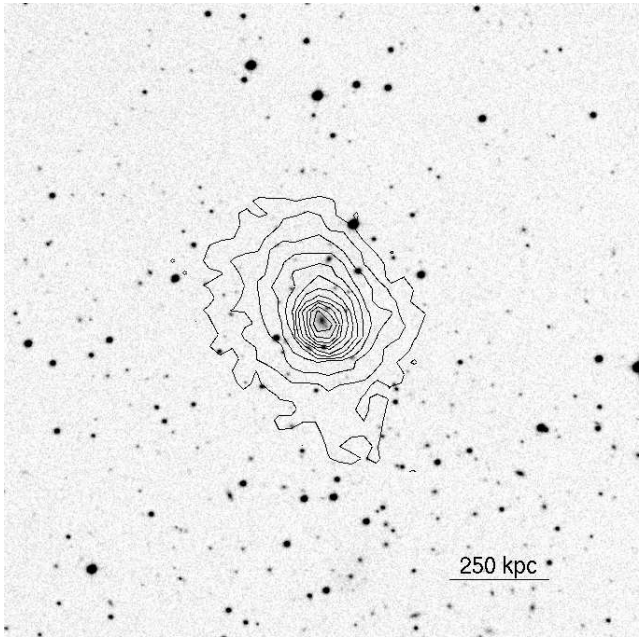}
\caption{DSS R-band image of A1664.  The X-ray contours show the emission spanning the entire ACIS-S3 chip.  North is toward the top; East is to the left.}
\end{figure}

\subsubsection{Surface Brightness Analysis}

We determined the center, ellipticity and position angle of A1664 on the sky by fitting a two-dimensional Lorentzian surface \citep{wise04}.  The centroid is found at R.A. =  $13^h03^m42^s.5$, DEC = $-24^{\degr}14'43''.96$.  This centroid is offset from the cluster emission peak by 3.6$''$ to the NE.  The average ellipticity was found to be 0.22 with a position angle of 25\degr.  Both are consistent with values found by \citet{all95}.  When dividing the data into annular regions for measuring radial trends, the centroid, ellipticity, and postion angle derived here were used.

An exposure map created using {\it mkexpmap} was divided into the image.  The surface brightness was calculated by measuring the flux through annuli 5$''$ in width.  We fit the surface brightness profile with a beta model using the Sherpa model {\it beta1d}.  A single beta model does not adequately fit the cluster surface brightness.  The model adequately fits the outer regions of the profile, but not the elevated brightness of the core.  Adding a second component to fit the cooling cusp improves the fit.  The form for the double-beta model is
\begin{equation}
I(r) = I_{1}(1+\frac{r^2}{r_{1}^2})^{-3\beta_{1}+\onehalf} + I_{2}(1+\frac{r^2}{r_{2}^2})^{-3\beta_{2}+\onehalf} + I_B,
\end{equation}
where $I_{i}$ and $r_{i}$ are the amplitudes and core radii.  The constant compontent, I$_B$, was added to represent the background.  All parameters from the fit can be found in Table 1.  The surface brightness profile presented in Figure 3 includes the fit from the double-beta model.  The smooth model does not fit the central region well, which is most likely due to the complicated structure of the core.

The residual map formed by subtracting the double-beta model is presented in Figure 4.  A spiral pattern is apparent near the center.  Similar spiral patterns have been seen in other clusters \citep{san02,cla04}.  The spiral is commonly attributed to the angular momentum of stripped gas from galaxies or groups.  Other explanations involve merging dark matter halos with trapped gas \citep{cla04,dup07}.

The spiral feature is common in systems with cold fronts \citep[see][for a review]{mar07}.  Most cold fronts are likely due to oscillations of the dark matter or gas in the cluster core \citep{tit05}.  This ``sloshing" can be triggered by a tidal or pressure disturbance caused by an infalling subcluster or possibly a powerful AGN outburst.  The sloshing gas can acquire angular momentum, which is why the spiral feature is associated with many cold fronts \citep{asc06}.  We present evidence that A1664 contains a cold front in section 3.2.5.

\begin{figure*}
\begin{center}
$\begin{array}{cc}
\leavevmode \epsfysize=8cm \epsfbox {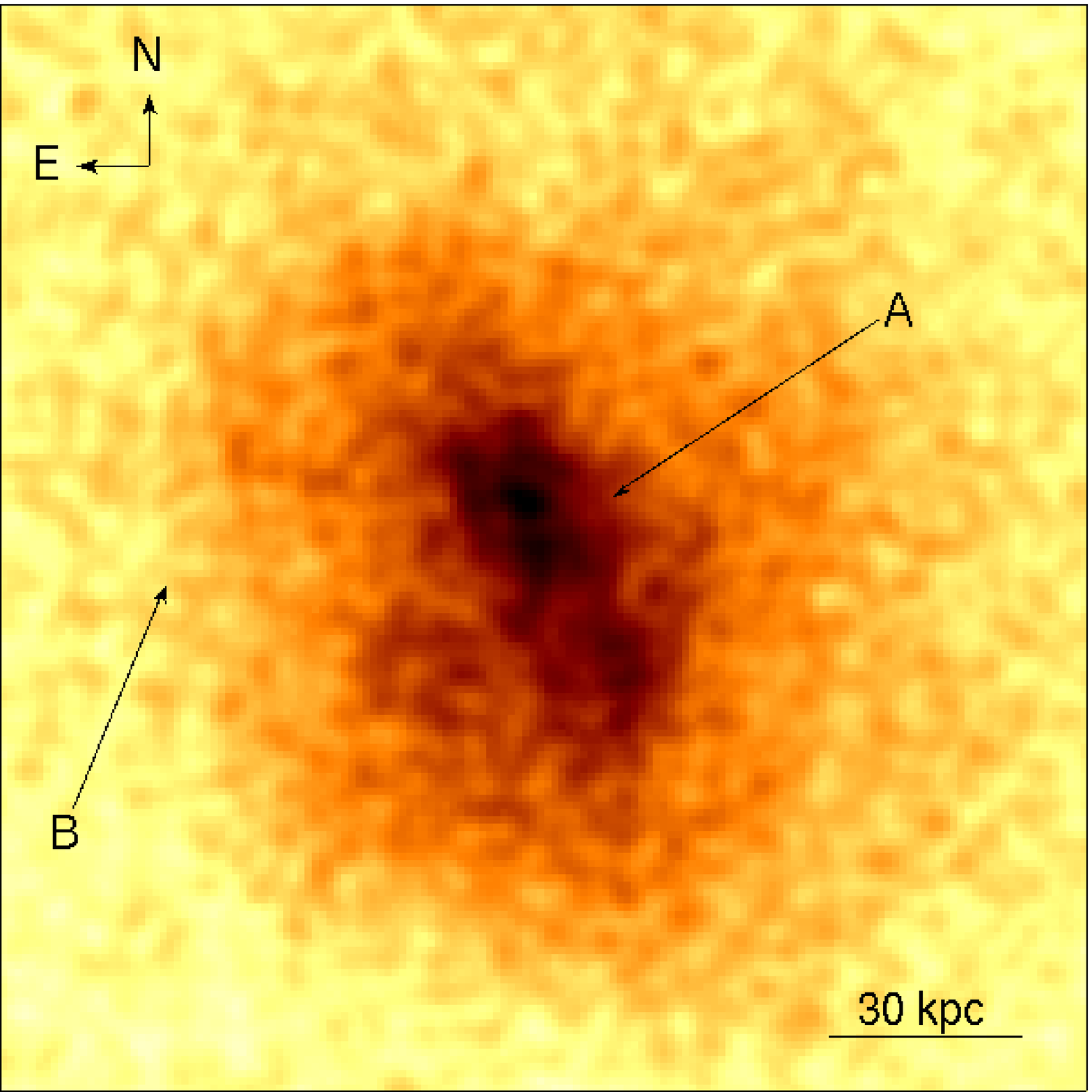} &
\leavevmode \epsfysize=8cm \epsfbox {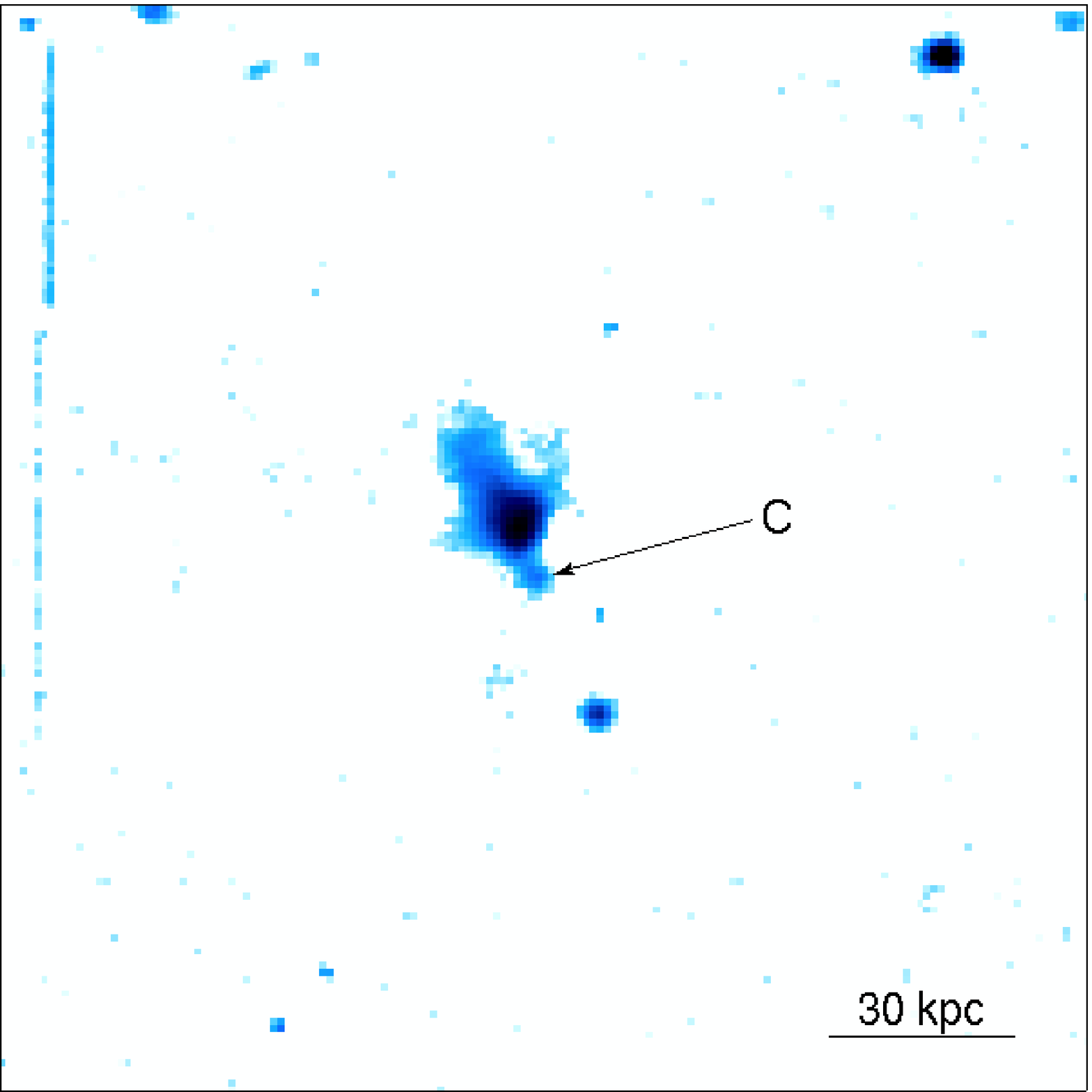} \\
\leavevmode \epsfysize=8cm \epsfbox {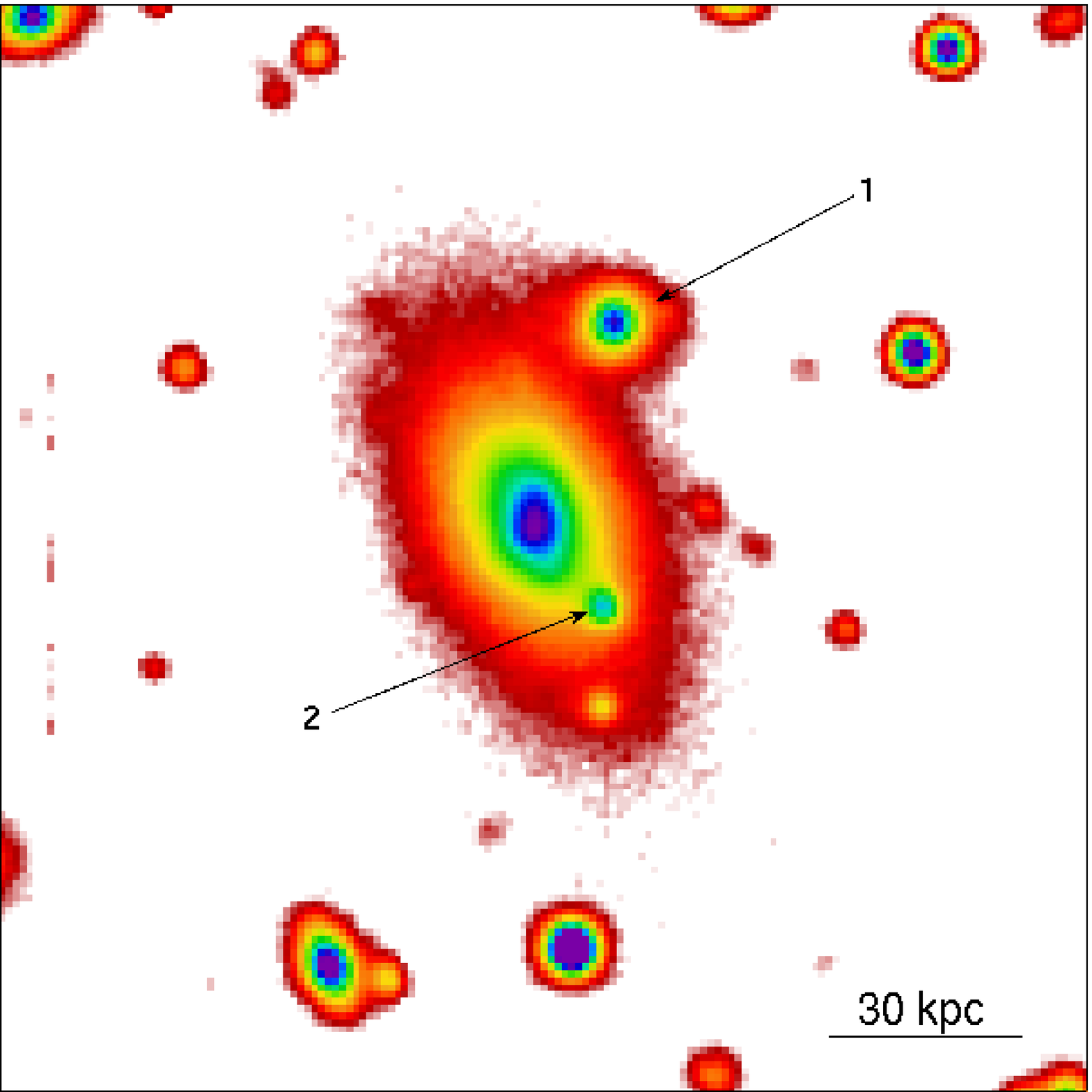} &
\leavevmode \epsfysize=8cm \epsfbox {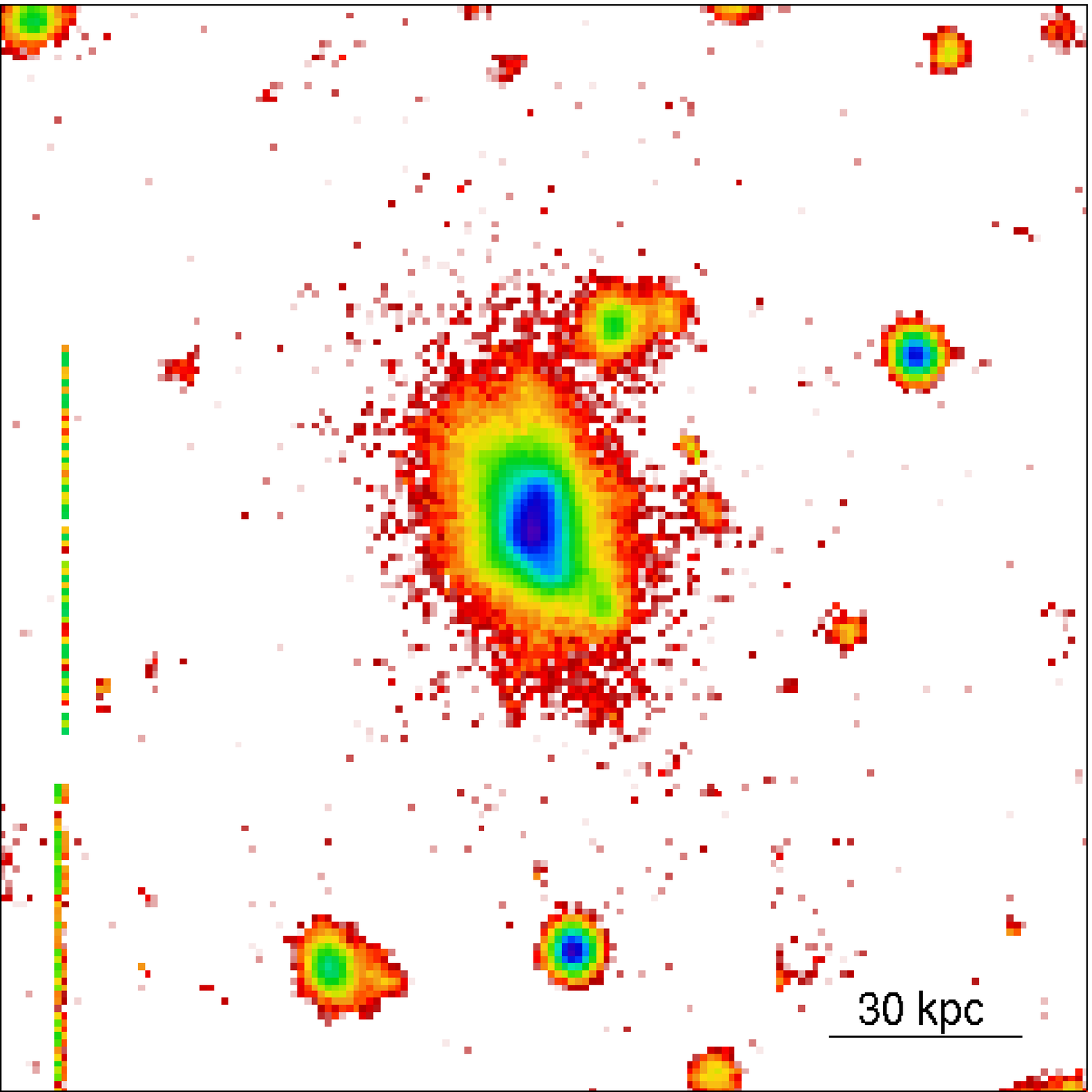} \\
\end{array}
$
\caption{{\it Upper left}: Inner 74$''$$\times$74$''$ region in the 0.4 - 7.0 keV band adaptively smooth using CIAO tool {\it aconvolve}.  {\it Upper right}: H$\alpha$ map of the same region.  {\it Lower left}: $R$-band image of the same region in false color to show structural detail.  {\it Lower right}:  $U$-band image of the same region in false color.  The labels are as follow: Point A is the X-ray bar, point B is the surface brightness edge, and point C is H$\alpha$ disruption being caused by a neighboring galaxy.  Points 1 and 2 are clusters members that may be interacting with the BCG.}
\end{center}
\end{figure*}

\subsubsection{Integrated Spectral Analysis}

For our spectral analysis, we begin by extracting the total spectrum of the cluster core using {\it dmextract}.  The elliptical region used in the extraction was chosen with semi-major axis of 161$''$ to stay within the edge of the ACIS-S3 CCD's footprint.  Weighted response files were created for the extracted regions using {\it mkacisrmf} and {\it mkwarf}.  Four different models were used to fit the properties of the X-ray gas over the energy range 0.4 keV to 7 keV with the data grouped with a minimum of 20 counts per bin.  Each model type was fit twice, once with the column density ($N_{H}$) as a free parameter and once with column density set at the Galactic value of 8.95$\times$10$^{20}$ cm$^{-2}$ \citep{dic90}.  All models allowed temperature, abundance, and normalization vary.  The simplest model used was a single temperature (1T) plasma with absorption (WABS$\times$MEKAL).  To test for multi-phase gas, a two temperature (2T) model was used by adding a second MEKAL component to the 1T model while tying the abundance parameters together.

We have also tested whether a cooling flow fits the spectrum by adding a cooling flow component to the single temperature model (WABS$\times$(MEKAL+MKCFLOW)).  The temperature of the thermal component was tied to the upper temperature of the cooling flow component.  Two separate fits were found by testing different scenarios based on the MKCFLOW model temperatures.  For the first scenario, we allowed both upper and lower temperatures to vary (VC), as well as abundance and normalization.  In the second case, we set the lower temperature, $kT_{low}$, to 0.1 keV \citep[e.g.,][]{wise04}, allowing the gas to cool fully (FC), i.e., to below the temperature range detectable by Chandra.  From this we found an upper limit on the mass condensation rate for the cluster within in the ACIS-S3 chip to be 137 $M_\sun$ yr$^{-1}$.  

The results are in Table 2.  The fits found for each model range in reduced $\chi^2$ values from 1.1 to 1.5.  Multiple temperature models are favored over the single temperature model, which is expected in a cooling-flow cluster.  The temperature variations will be examined more closely in section 3.2.5.  All models with $N_{H}$ allowed to vary proved better fits than the models set to the Galactic value, suggesting an error in the interpolation of the foreground value or error in the calculation of the buildup of hydrocarbons on the optical filter.  The latter possibility is unlikely.  A detailed CXC memo by Vikhlinin\footnote{See \url{http://hea-www.harvard.edu/$\sim$alexey/acis/memos/}.} claims the calibration of the buildup is known to less than 5\% at low energies.

To understand how this absorption is distributed across the field of view, we created elliptical annuli containing about 5000 net counts.  Fitting each of these regions with a single temperature model over the same energy range as before, we produced the $N_{H}$ radial profile shown in Figure 5.  The dotted line indicates the Galactic value for hydrogen column density.  The profile shows no radial trend, but overall the column density is above the Galactic value in all regions.  The average column density for this region is better fit with the value 1.22$\times$10$^{21}$ cm$^{-2}$.  This value is in agreement with the fitted values for the integrated spectrum.  For all models used in this analysis, a significantly better fit is found using this average column density rather than the Galactic value.  Therefore, we have adopted it for the remainder of this paper.

\begin{deluxetable*}{cccccccccc}
\tablecolumns{10}
\tablewidth{0pc}
\tablecaption{Beta-Model Fit}
\tablehead{
	\colhead{$r_1$} & \colhead{} & \colhead{} & \colhead{$r_2$} & \colhead{} & \colhead{} & \colhead{} & \colhead{} & \colhead{} & \colhead{} \\
	\colhead{(arcsec)} & \colhead{$\beta_1$} & \colhead{$I_1$\tablenotemark{a}} & \colhead{(arcsec)} & \colhead{$\beta_2$} & \colhead{$I_2$\tablenotemark{a}} & \colhead{$I_B$\tablenotemark{a}} & \colhead{$\chi_{norm}^2$} & \colhead{$\chi^2$} & \colhead{$d.o.f.$}
	}
\startdata
	$89.83_{-2.36}^{+2.39}$ & $0.78_{-0.02}^{+0.02}$ & $75.7_{-1.4}^{+1.3}$ & $23.65_{-0.36}^{+0.34}$ & $0.91_{-0.01}^{+0.01}$ & $1220_{-17}^{+16}$ & $6.93_{-0.09}^{+0.09}$ & 2.57 & 228.48 & 89 \\
\enddata
\tablenotetext{a}{Units of 10$^{-9}$ phot sec$^{-1}$ cm$^{-2}$ arcsec$^{-2}$}
\end{deluxetable*}

\begin{figure}[b]
\epsscale{1.25}
\plotone{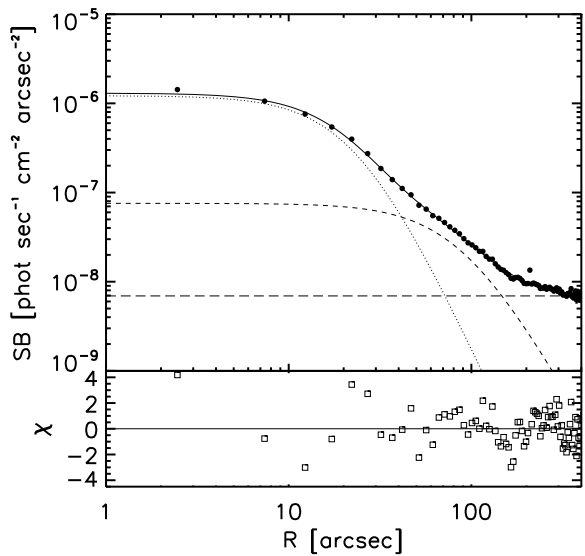}
\caption{Radial surface brightness profile in the 0.4 - 7.0 keV band.  The fit shown represents the two components of the double-beta model and the best fit background value.  The panel below shows the residual deviation from the fit.}
\end{figure}

\subsubsection{The X-ray Bar}

The bar region, indicated in Figure 2 (A), stands out as a strong peak in surface brightness.  We determined its dimensions from its average surface brightness profiles in the directions perpendicular and parallel to its major axis.  Using the full-width at half-maximum of the average peaks found in Figure 6, the length is 22.5$''$ and width is 10$''$.  These dimensions were used to isolate its spectrum.  A total of 3350 net counts were detected in the extracted region.  The center of a cluster suffers the most from projection effects, so contributions from four annular regions outside of the bar containing 15,000 net counts each were fitted for deprojection.  The resulting fit of the PROJCT(WABS$\times$MEKAL) model for the bar is presented in Table 3.  The electron density of the region is 0.066 cm$^{-3}$, corresponding to a total mass of 3.21$\times$10$^{10}$ $M_\sun$.  If the gas in the bar was all cooling to low temperatures on its cooling timescale, it would be depositing about 91 $M_\sun$ yr$^{-1}$.  The density, pressure, entropy, and cooling time are all consistent with the values fit for the central region of the radial analysis, found later in section 3.2.5.

\begin{figure}[b]
\epsscale{1.15}
\plotone{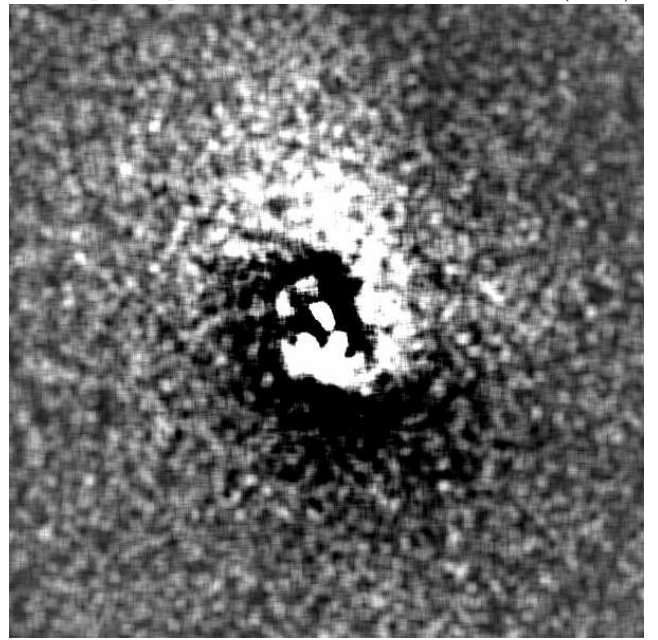}
\caption{A 5$'$$\times$5$'$ residual map of the double-beta model subtracted image.  The image was adaptively smooth using CIAO tool {\it aconvolve} before subtraction.}
\end{figure}

We will now consider the possibility that the bar is actually shaped more like an edge-on disk.  Before we assumed that the bar is prolate, with its major axis in the plane of the sky.  The edge-on disk is oblate, with its minor axis assumed to lie in the plane of the sky.  With a larger volume because of the oblate geometry, a lower density will be measured.  These differences are presented in Table 3.  With these assumptions the density and pressure decrease to 0.0443 cm$^{-3}$ and 2.42$\times$10$^{-10}$ erg cm$^{-3}$, respectively.  The entropy and cooling time rise to 13.6 keV cm$^2$ and 5.3$\times$10$^8$ yr, respectively.  The mass increases to 7.22$\times$10$^{10}$ $M_\sun$, more than double the mass contained within the bar geometry.  The assumed geometry does not affect our conclusions.

The CO-derived molecular gas mass for A1664 is 4.4$\times$10$^{10}$ $M_\sun$ \citep{edge01}.  The total mass of the X-ray emitting gas in the bar is comparable to the cold molecular gas that resides at the same location.  A study by \citet{wil06} has shown that the CO and H$\alpha$ emission are tracing the same gas, which is shown in the upper right panel of Figure 2.  The X-ray bar structure is associated with this same disturbed region.

\begin{deluxetable*}{lcccccc}
\tablecolumns{7}
\tablewidth{0pc}
\tablecaption{Integrated Spectral Fits}
\tablehead{
	\colhead{} & \colhead{$kT$} & \colhead{$kT_{low}$} & \colhead{} & \colhead{$N_{H}$} & \colhead{} & \colhead{} \\
	\colhead{Model} & \colhead{(keV)} & \colhead{(keV)} & \colhead{$Z$} & \colhead{$(10^{22} cm^{-2})$} & \colhead{$\dot{M}_{X}$} & \colhead{$\chi^2 / d.o.f$}
	}
\startdata
	1T & $3.65_{-0.05}^{+0.05}$ & --- & $0.46_{-0.03}^{+0.03}$ & $0.116_{-0.003}^{+0.003}$ & --- & $481.26/368$ \\
	1T & $4.11_{-0.06}^{+0.06}$ & --- & $0.54_{-0.03}^{+0.03}$ & $0.0895$ & --- & $564.01/369$ \\
	2T & $5.17_{-0.32}^{+0.87}$ & $1.69_{-0.10}^{+0.12}$ & $0.37_{-0.04}^{+0.03}$ & $0.120_{-0.003}^{+0.004}$ & --- & $402.25/366$ \\
	2T & $5.25_{-0.36}^{+0.34}$ & $1.65_{-0.22}^{+0.11}$ & $0.49_{-0.03}^{+0.03}$ & $0.0895$ & --- & $503.93/367$ \\
	VC & $4.31_{-0.11}^{+0.11}$ & $0.37_{-0.04}^{+0.04}$ & $0.46_{-0.03}^{+0.03}$ & $0.129_{-0.004}^{+0.003}$ & $145_{-18}^{+16}$ & $397.59/366$ \\
	VC & $8.53_{-3.01}^{+0.52}$ & $1.33_{-0.09}^{+0.08}$ & $0.51_{-0.03}^{+0.03}$ & $0.0895$ & $367_{-83}^{+21}$ & $503.19/367$ \\
	FC & $4.23_{-0.11}^{+0.11}$ & $0.1$ & $0.47_{-0.03}^{+0.03}$ & $0.132_{-0.004}^{+0.004}$ & $137_{-17}^{+17}$ & $400.63/367$ \\
	FC & $4.30_{-0.09}^{+0.09}$ & $0.1$ & $0.56_{-0.03}^{+0.03}$ & $0.0895$ & $29_{-10}^{+10}$ & $556.21/368$ \\
\enddata
\end{deluxetable*}

\begin{figure}[b]
\epsscale{1.25}
\plotone{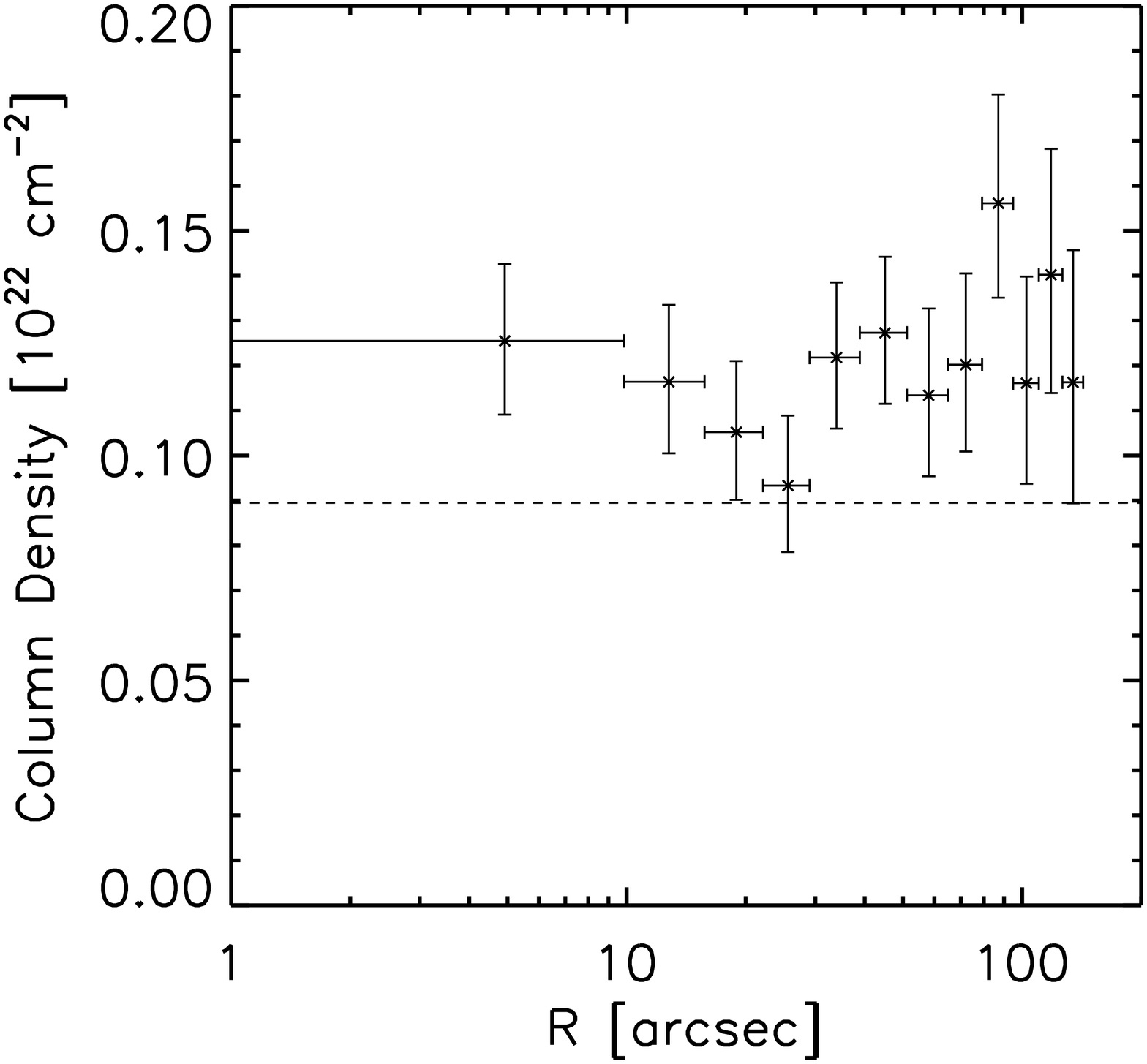}
\caption{Hydrogen column density profile (projected).  The dashed line represents the Galactic value.}
\end{figure}

\subsubsection{Metal Abundances}

The metallicity analysis was carried out by creating adaptively sized regions containing an average 5000 net counts out to the edge of the CCD (161$''$).  All regions have their own weighted response files created using {\it mkacisrmf} and {\it mkwarf}.  After the spectrum for each region was extracted, each was modeled as a single temperature plasma (MEKAL) with an absorption component (WABS).  The model was fit over the energy range 0.4 to 7.0 keV, with $N_{H}$ set to our adopted value.  Temperature, metallicity, and normalization were allowed to vary.  The abundance ratios used in this analysis are from the most current photospheric data of \citet{gre98}.

The projected metallicity profile is presented in the left panel of Figure 7 represented by the solid points.  The abundance rises from the center of the cluster at 0.54 solar to 0.75 solar at 21$''$ and remains constant out to about 45$''$.  The metallicity then declines to approximately 1/3 of the solar value at a radius of 100$''$.  The central dip in metallicity is unusual, although central abundance dips have been observed in several clusters such as Centaurus A \citep{san02}, Perseus \citep{sch02}, and M 87 \citep{boh01}.  Usually the metallicity rises to the center in cooling flows \citep{deg01}.  Our current understanding of the central metal enhancement of cooling flow clusters posits that SN Ia in the central galaxy are enriching the gas with iron and other metals.  The peak metallicity is expected to be centered on the BCG.  These rare systems with central dips may be explained by resonance scattering \citep{gil87}, multiphase gas \citep{buo00}, warm+hot ISM \citep{buo01}, or bimodal metallicity distributions \citep{mor03}.

We explore the metallicity profile first by taking into account projection effects.  A deprojected fit was carried out using 5 radial bins, containing roughly 10000 net counts each.  The spectrum for each region was modeled as a single temperature plasma with foreground absorption, but now with the additional projected component.  The open circles in Figure 7 representing the deprojected profile show that the abundance dip becomes more pronounced.  Overall the 5 bins generally remain within errors of the projected profile.  The peak at 45$''$ lies at a value of 0.93 solar.  To within errors, this value is consistent with the projected value at the same radius.

We now explore the  possibility of multiphase gas and low energy absorption due to warm gas.  \citet{mol01} have shown that a single temperature model can bias a measurement to lower metallicities.  It was previously demonstrated by \citet{buo00} that this bias is due to the abundance measurement being sensitive to the Fe-L complex at low temperatures.  In a multiphase medium, the Fe-L complex is excited by a range of temperatures around 1 keV, resulting in a broader spectral shape.  A single, average temperature would treat this broad peak as a lower metallicity feature, resulting in an underestimation of the actual abundance.

Using the same regions and model we used for the projected analysis, we added a cooling flow component to the model (WABS$\times$(MEKAL+MKCFLOW)) with the upper temperature tied to the temperature of the thermal component.  The lower temperature was set at 0.1 keV to allow full cooling.  A significant improvement in the fit was found for the inner two regions when adding this component.  An F-test shows that the outer six regions improve slightly or remain the same.  The reduced chi-squared values for both models are presented in Table 4.

\begin{figure*}[t]
\begin{center}
$\begin{array}{cc}
\leavevmode \epsfysize=8cm \epsfbox {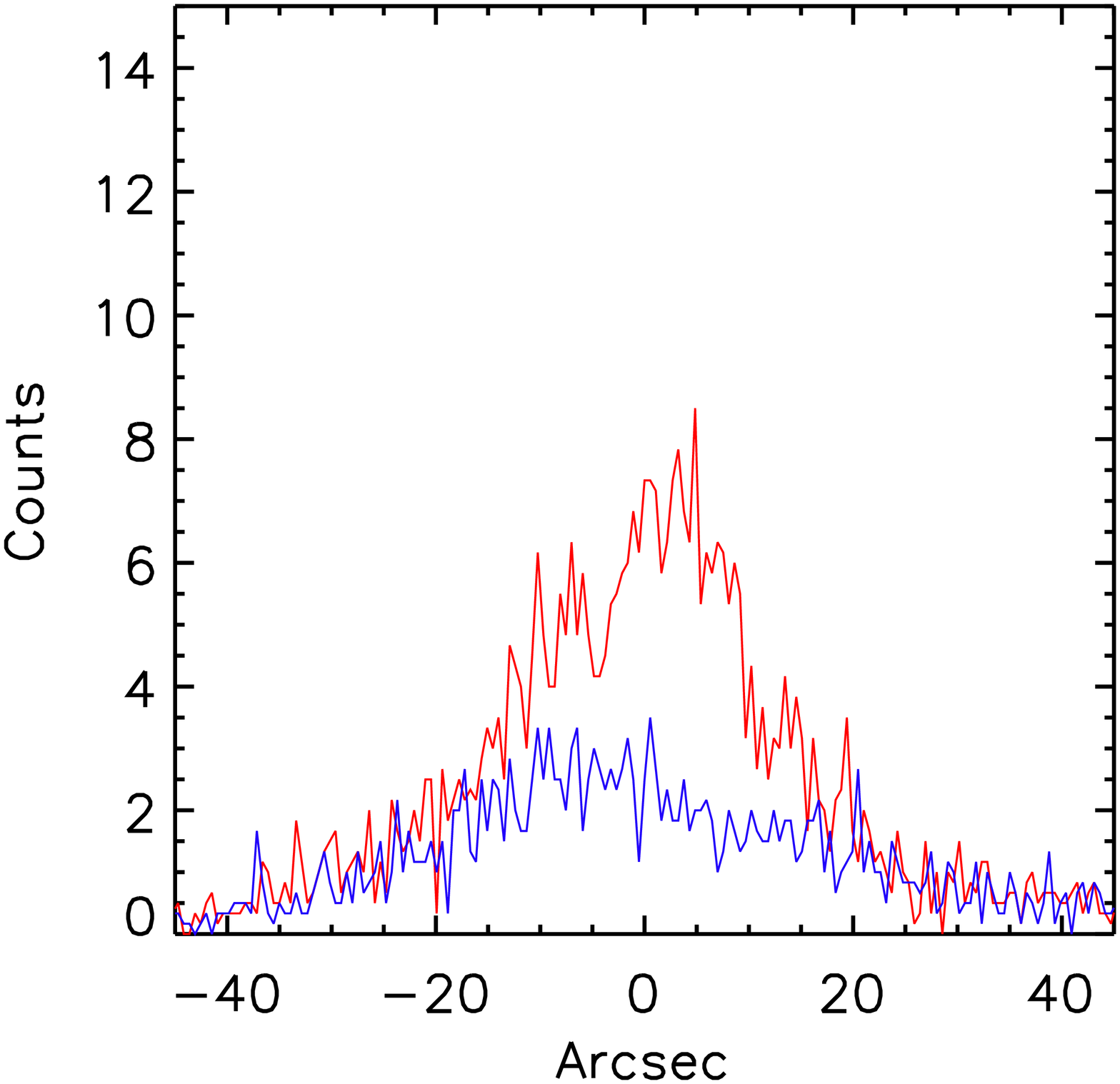} &
\leavevmode \epsfysize=8cm \epsfbox {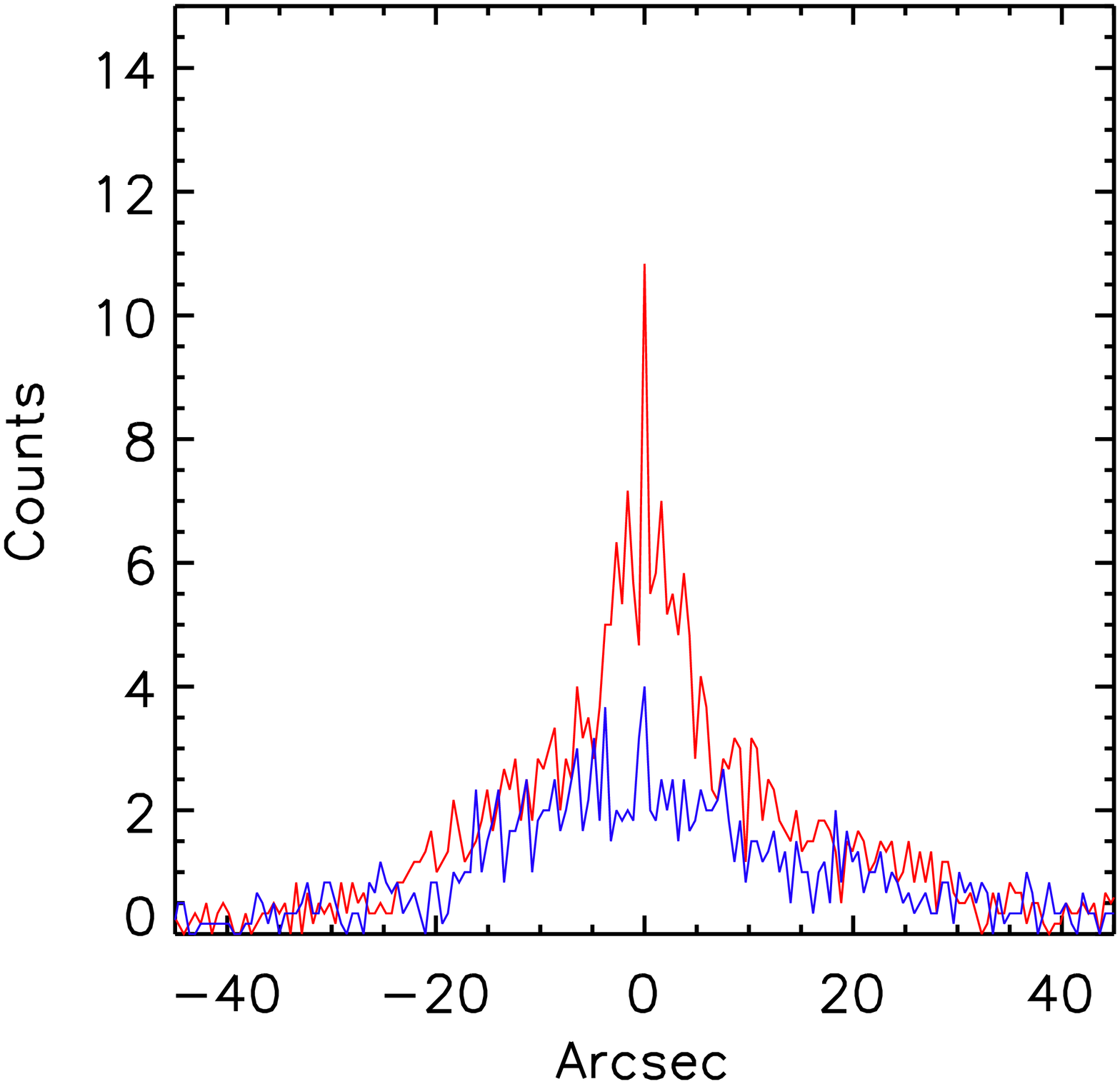} \\
\end{array}
$
\caption{{\it Left}: Average surface brightness (in total counts) parallel to the length of the bar.  {\it Right}: Average surface brightness (in total counts) parallel to the width of the bar.  The red lines indicates the profile directly through the bar.  The blue line is the average profile of the region immediately outside of the bar for comparison.}
\end{center}
\end{figure*}

\begin{deluxetable*}{ccccccccc}
\tablecolumns{7}
\tablewidth{0pc}
\tablecaption{Spectral Fits for Core Region}
\tablehead{
	\colhead{} & \colhead{$n_e$} & \colhead{$P$} & \colhead{$S$} & \colhead{$t_{cool}$} & \colhead{$M$} & \colhead{} \\
	\colhead{Geometry} & \colhead{($10^{-2}$ cm$^{-3}$)} & \colhead{($10^{-10}$ erg cm$^{-3}$)} & \colhead{(keV cm$^2$)} & \colhead{($10^8$ yrs)} & \colhead{($10^{10}$ $M_\sun$)} & \colhead{$\dot{M}_{X}$}
	}
\startdata
	Bar & 6.64$_{-0.28}^{+0.27}$ & 3.63$_{-0.23}^{+0.21}$ & 10.4$_{-0.6}^{+0.5}$ & 3.5$_{-0.4}^{+0.4}$ & 3.21$_{-0.14}^{+0.12}$ & 91$_{-11}^{+11}$ \\
	Disk & 4.43$_{-0.19}^{+0.17}$ & 2.42$_{-0.15}^{+0.14}$ & 13.6$_{-0.7}^{+0.7}$ & 5.3$_{-0.6}^{+0.6}$ & 7.22$_{-0.30}^{+0.28}$ & 140$_{-16}^{+16}$ \\
\enddata
\end{deluxetable*}

The new abundance profile with the inner two regions modeled with cooling flows is shown in the right panel of Figure 7.  The central metallicity dip has largely gone away.  To within the errors, the profile is centrally peaked.  The upper limit on the mass condensation rate for these two regions was found to be 56$\pm$10 $M_\sun$ yr$^{-1}$.  In an attempt to improve the fit of the inner most region further, we assume that there may be reservoir of warm gas ($\sim$ 10$^5$ K), being deposited from the cooling flow, acting as a source of extra absorption in the lower energy range of the spectrum (0.4 - 0.8 keV).  There is a known reservoir of cold gas that is comparable in mass to the hot X-ray emitting gas associated with the BCG (discussed in sec 3.3.1).  Our assumption is that if the cooling flow is directly feeding the cold gas, then there would be gas in the intervening temperature range, though the amount expected is small due to extremely short cooling times and therefore the effect should be minimal.  This extra absorption was represented using the EDGE component in XSPEC.  A visual inspection of the spectrum at lower energies shows the fit is no longer systematically higher than the data, which increases the metallicity measurement to the solar value.  However, by introducing two new free parameters to the model, the chi-squared value did not improve significantly.

\begin{figure*}[t]
\begin{center}
$\begin{array}{cc}
\leavevmode \epsfysize=8cm \epsfbox {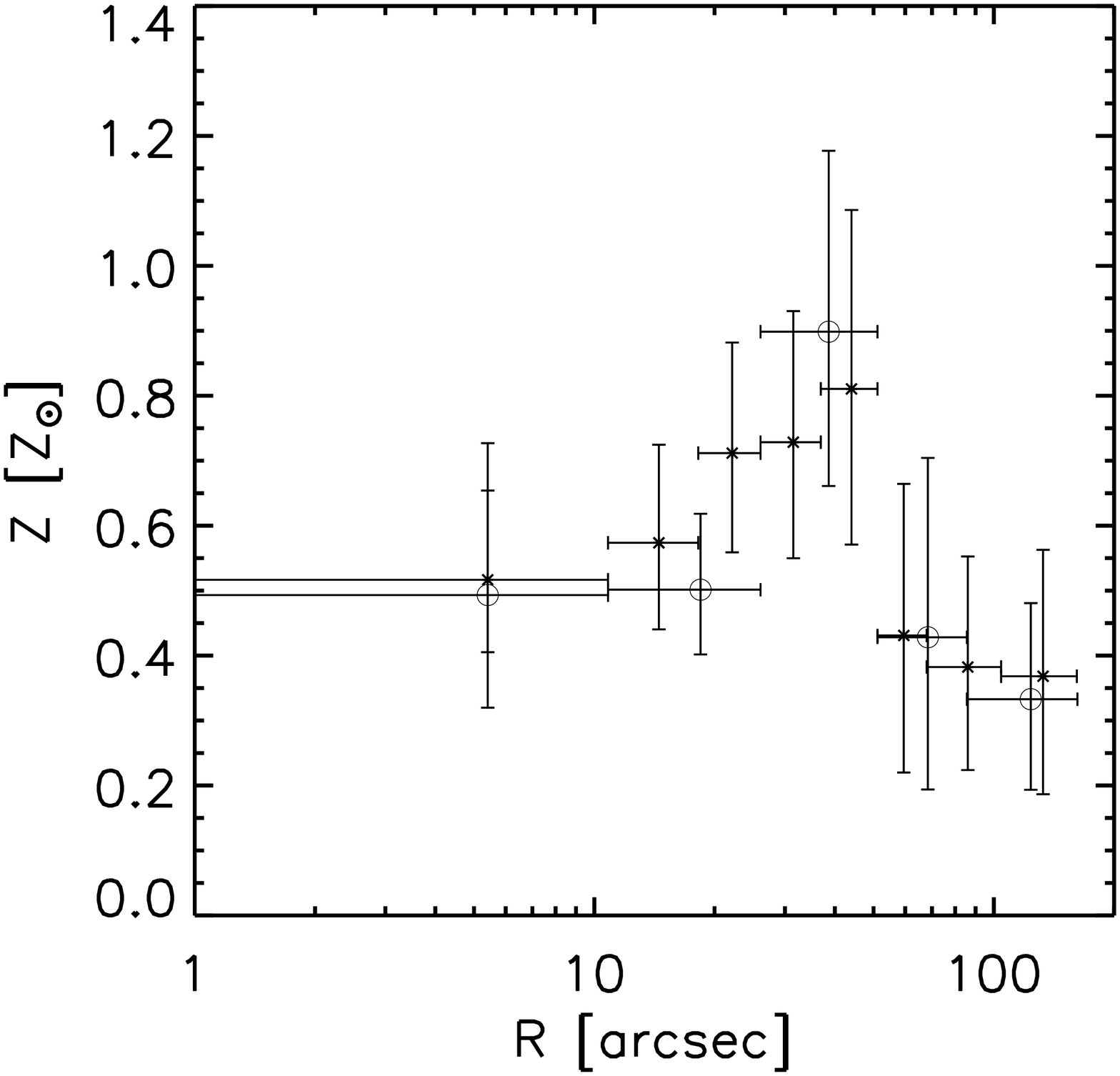} &
\leavevmode \epsfysize=8cm \epsfbox {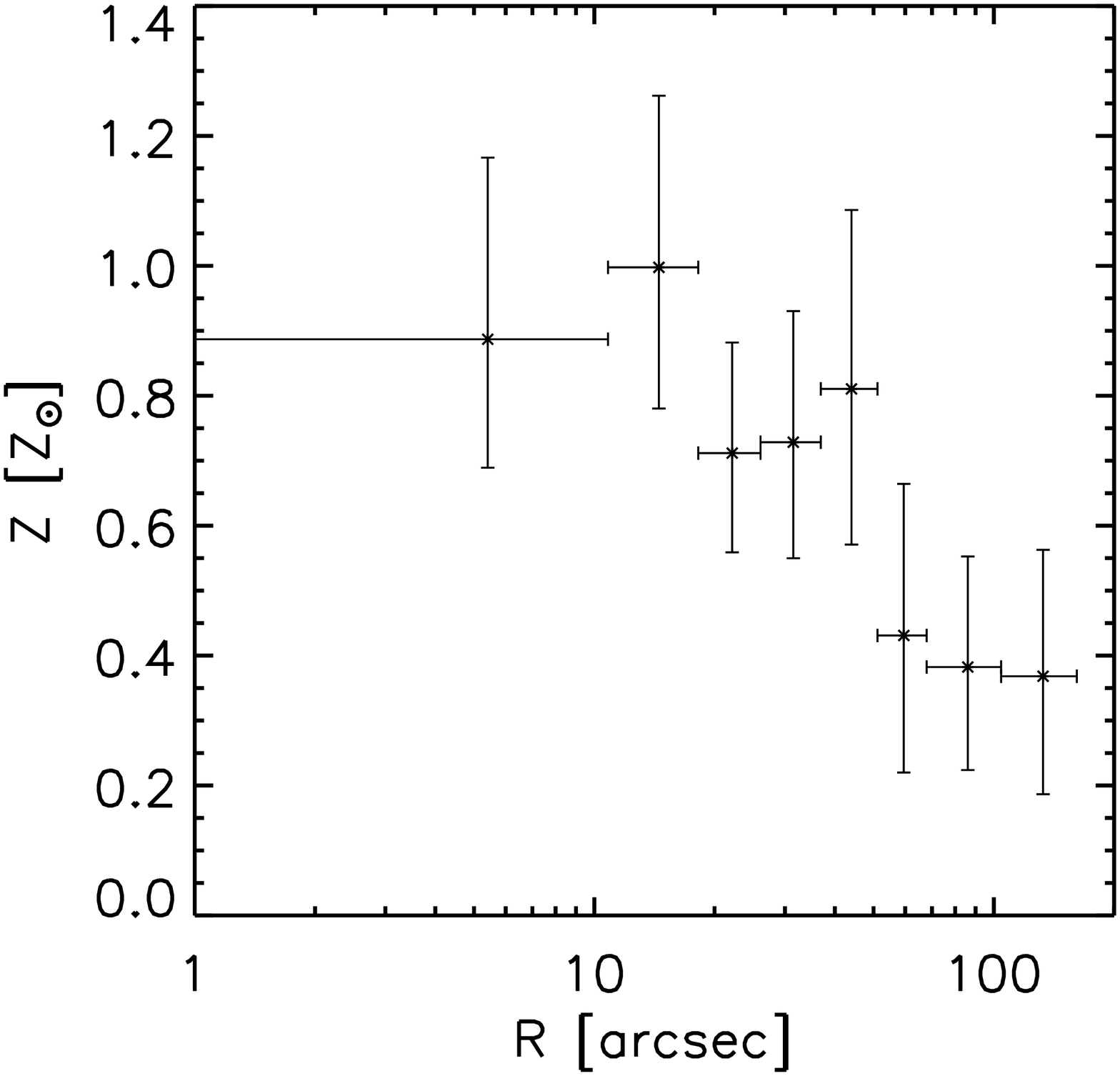} \\
\end{array}
$
\caption{{\it Left}: Projected ({\it asterisks}) and deprojected ({\it circles}) metallicity profile of the ICM in solar units using a single temperature model.  An unusual dip occurs in the central $\sim$ 50$''$.  {\it Right}: Metallicity profile of ICM in solar units.  The inner two points were derived by including a cooling flow model to obtain a better metallicity estimation.}
\end{center}
\end{figure*}

\subsubsection{Physical State of the Gas}

A radial temperature analysis was performed by extracting regions from the same area of the CCD as the metallicity analysis.  The sizes of the annuli were adjusted so that each contained approximately 2000 net counts.  The radial bins were fit in projection following the same procedure and using the same model as described above in the metallicity analysis. 

Shown in Figure 8 as asterisks, the temperature reaches a minimum of 2.2 keV at the center of the cluster.  At a radius of about 18$''$, the temperature rises from $2.6\pm0.1$ keV to $2.9\pm0.2$ keV and drops back down to $2.5\pm0.2$ keV in the next annulus.  This feature may be due to an asymmetric temperature structure caused by the eastern cold front.  From there, the profile rises steeply to 4.6 keV at about 45$''$ and reaches its peak temperature of 5.7 keV at around 100$''$.  We note that when fitting the temperatures with the abundances tied to those derived using the cooling flow model in section 3.2.4, the changes in the best fit values are imperceptible.

The deprojected temperature profile was divided into 8 elliptical annuli.  The temperature was determined using the PROJCT component in XSPEC, as before.  The open circles represent the deprojected temperature profile in Figure 8.  Apart from the central bins, the deprojected profile agrees with the projected profile.  The central temperature decreases to 1.7 keV, making it 0.5 keV cooler than found in the projected profile.  Between the second and third radial bin, the temperature jump is still apparent in the temperature structure.

Density, pressure, entropy, and cooling-time profiles have been derived from the same radial bins.  The density profile, obtained from the deprojection of the surface brightness profile, is shown in the upper left panel of Figure 9.  The core reaches an electron density of 0.07 cm$^{-3}$, which is comparable to that found in the core of the Perseus cluster \citep{fab06}.  The profile decreases radially to below 0.002 cm$^{-3}$ in the last annulus at a radius of 160$''$. 

The pressure profile was calculated as $2 kT n_e$.  The upper right panel of Figure 9 shows the pressure peaking at the center at a value of 5$\times$10$^{-10}$ erg cm$^{-3}$.  At 20$''$, there is a break in the profile.  This is related to the temperature decrease described above.

\begin{deluxetable*}{ccccccccc}
\tablecolumns{9}
\tablewidth{0pc}
\tablecaption{Reduced Chi-squared Fits for Radial Metallicity Profile}
\tablehead{
	\colhead{} & \multicolumn{8}{c}{Reduced chi-squared for region (d.o.f.)} \\
	\cline{2-9} \\
	\colhead{Model} & \colhead{Reg. 1} & \colhead{Reg. 2} & \colhead{Reg. 3} & \colhead{Reg. 4} & \colhead{Reg. 5} & \colhead{Reg. 6} & \colhead{Reg. 7} & \colhead{Reg. 8}
 	}
\startdata
	1T & 1.15(133) & 1.35(146) & 1.06(148) & 1.20(153) & 0.99(156) & 1.17(152) & 0.95(206) & 1.12(234) \\
	CF & 0.98(132) & 1.07(145) & 0.97(147) & 1.15(152) & 0.92(155) & 1.17(151) & 0.93(205) & 1.10(233) \\
\enddata
\end{deluxetable*}

The entropy profile in the lower left panel of Figure 9 was calculated using the standard relationship $S=kT n_e^{-2/3}$.  The profile shows evidence for feedback in the core.  A cluster without heating would have an entropy profile that decreases roughly as a power law to the center \citep{voit02,kay04}.  We find that the profile decreases inward and then flattens at small radii  around 13 keV cm$^2$.  This entropy floor is normally found in cooling flow clusters \citep{voit05} where AGN activity is thought to be flattening out the profile by periodically releasing energy back into the surrounding gas.  We present the cooling time as a function of radius in the lower right panel of Figure 9.  The short central cooling times are characteristic of a cooling flow cluster.  The decreasing cooling time towards the center reaches its minimum value of 3.6$\times$10$^8$ yr, which is consistent with star forming cooling flows.  The sample of \citet{raf08} showed that star formation is preferentially found in systems with a central cooling time less than approximately 5$\times$10$^8$ yr, and with entropy less than 30 keV cm$^2$.

\begin{figure}[b]
\epsscale{1.25}
\plotone{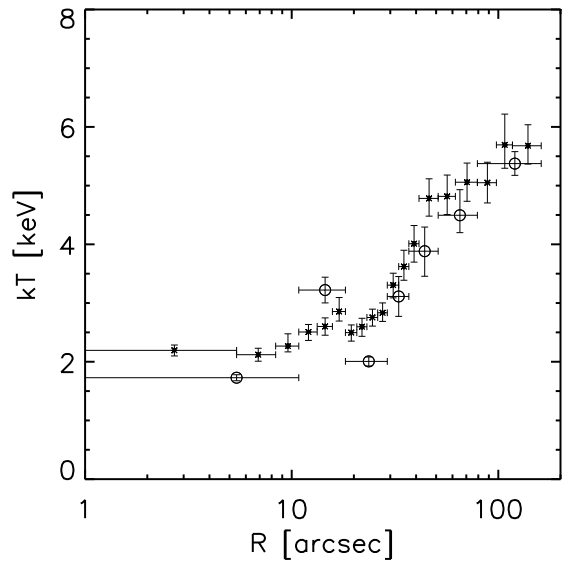}
\caption{Projected ({\it asterisks}) and deprojected ({\it circles}) X-ray temperature of the ICM.}
\end{figure}

\begin{figure*}[t]
\begin{center}
$\begin{array}{cc}
\leavevmode \epsfysize=8cm \epsfbox {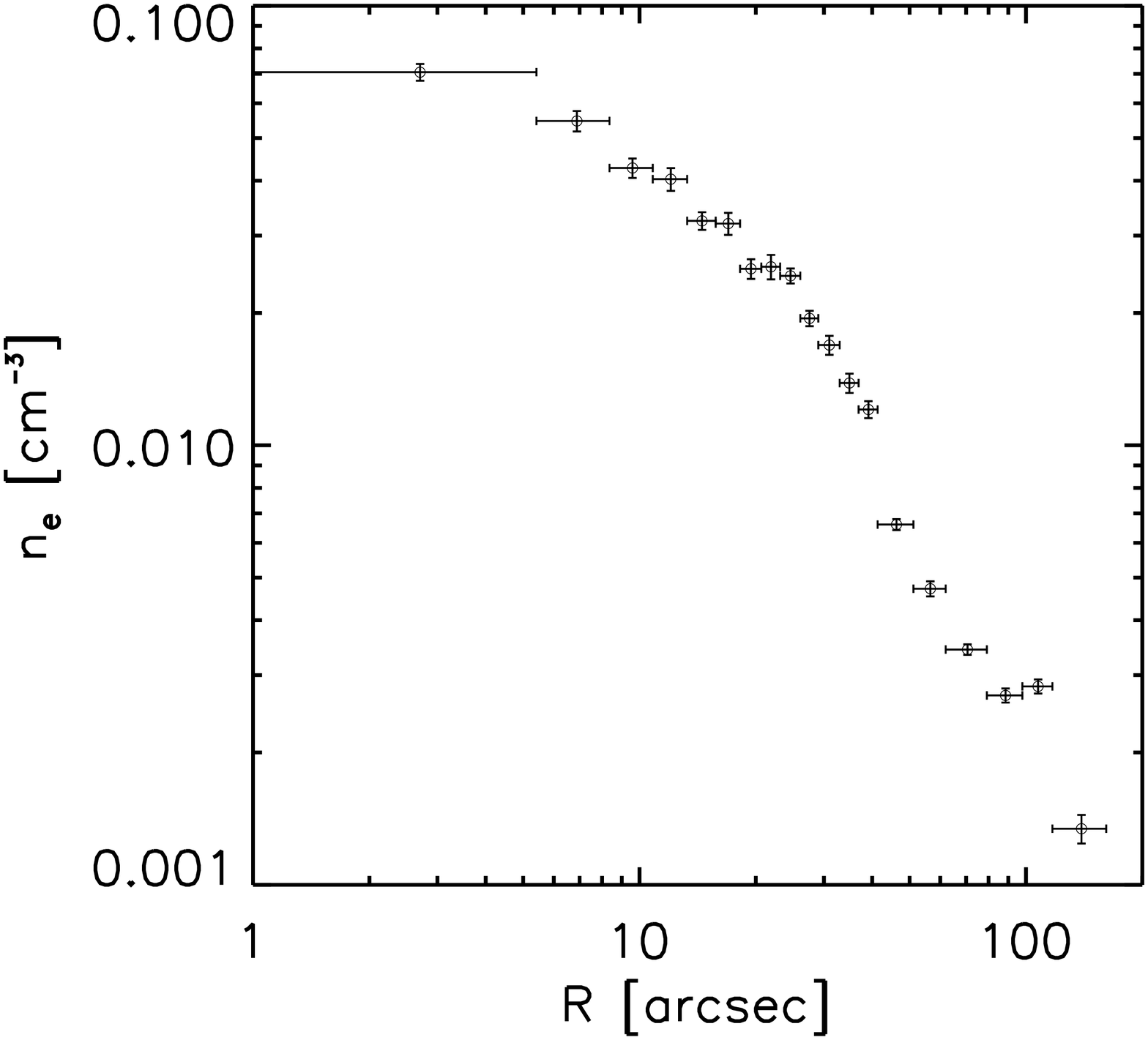} &
\leavevmode \epsfysize=8cm \epsfbox {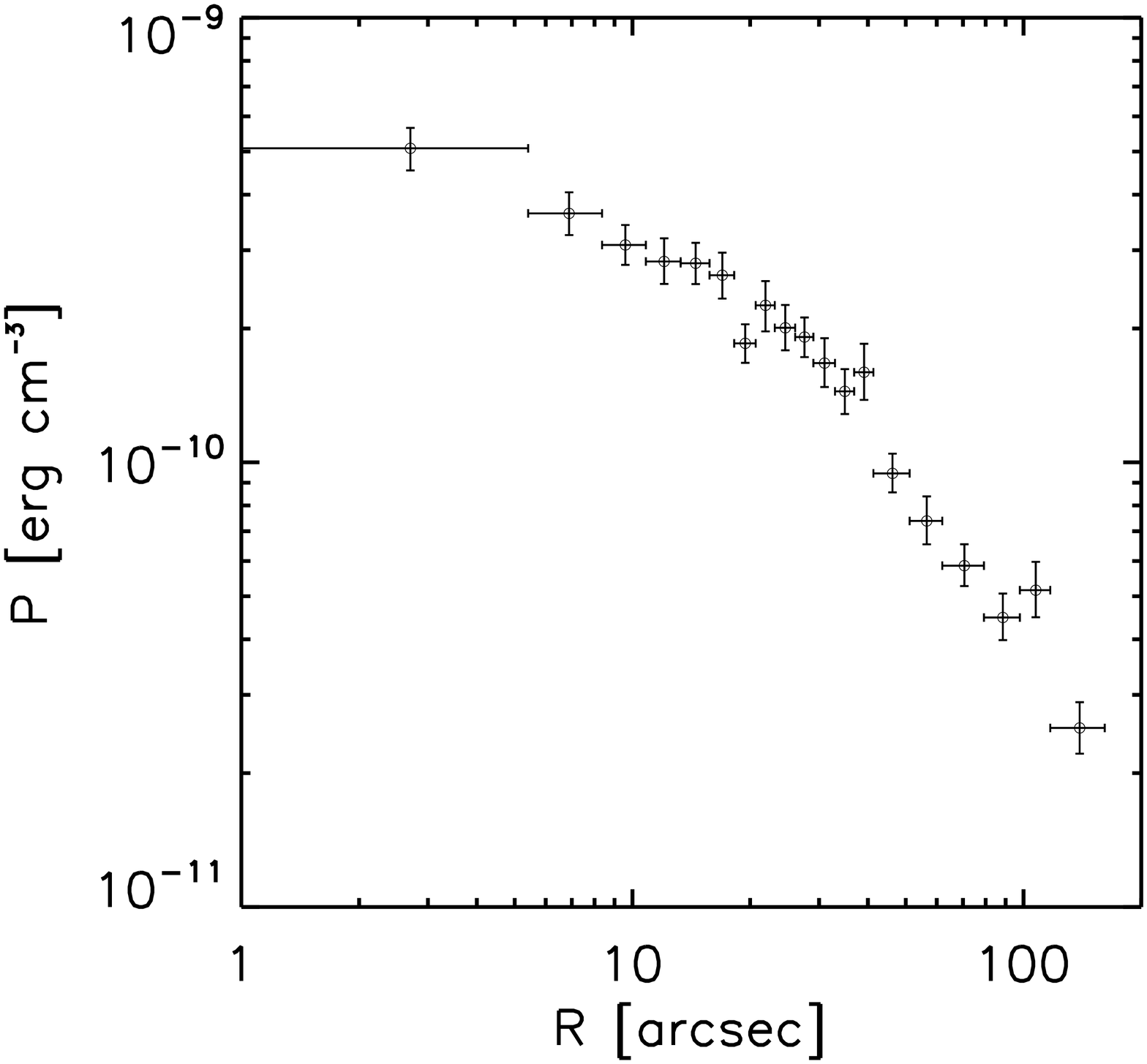} \\
\leavevmode \epsfysize=8cm \epsfbox {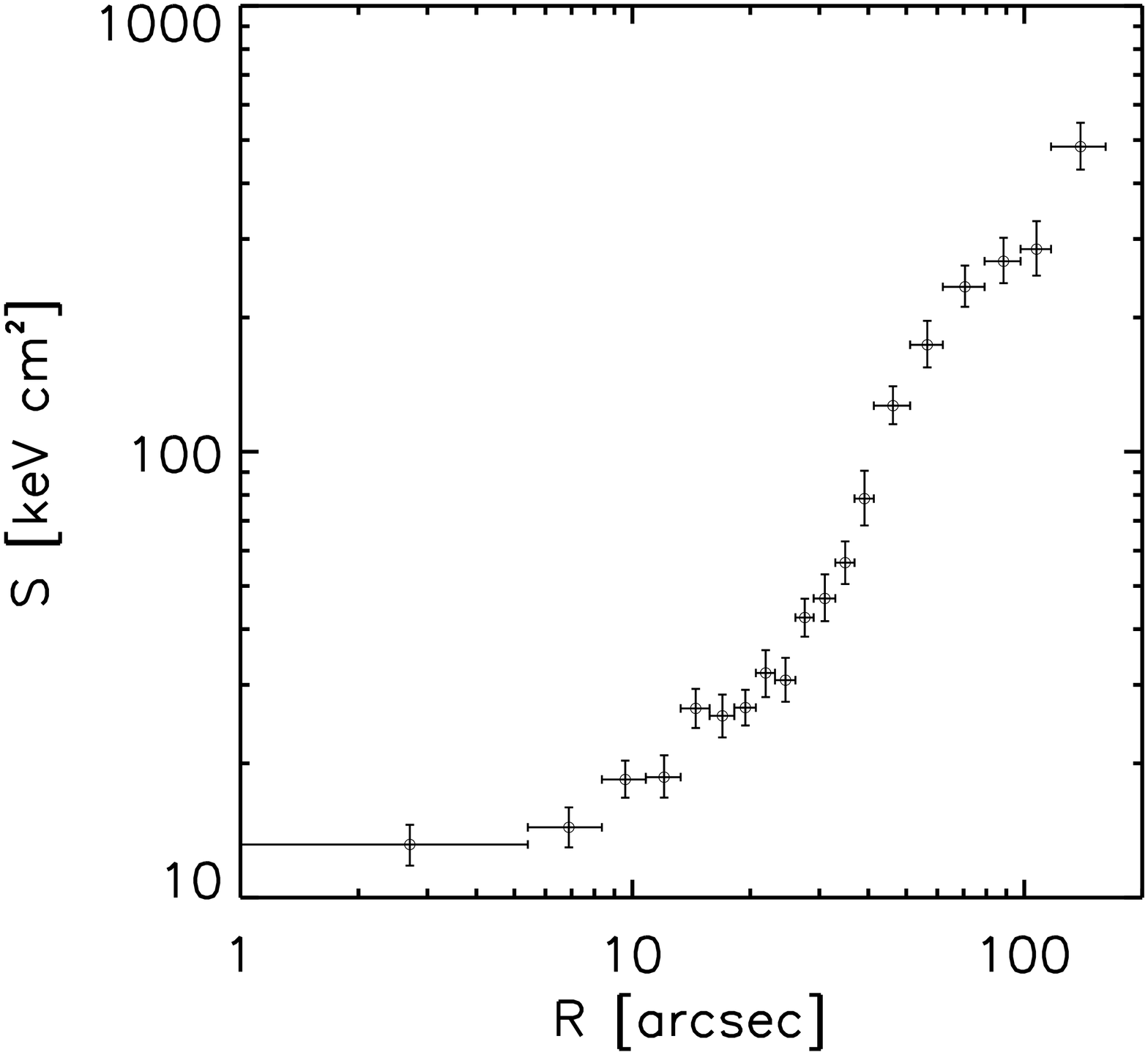} &
\leavevmode \epsfysize=8cm \epsfbox {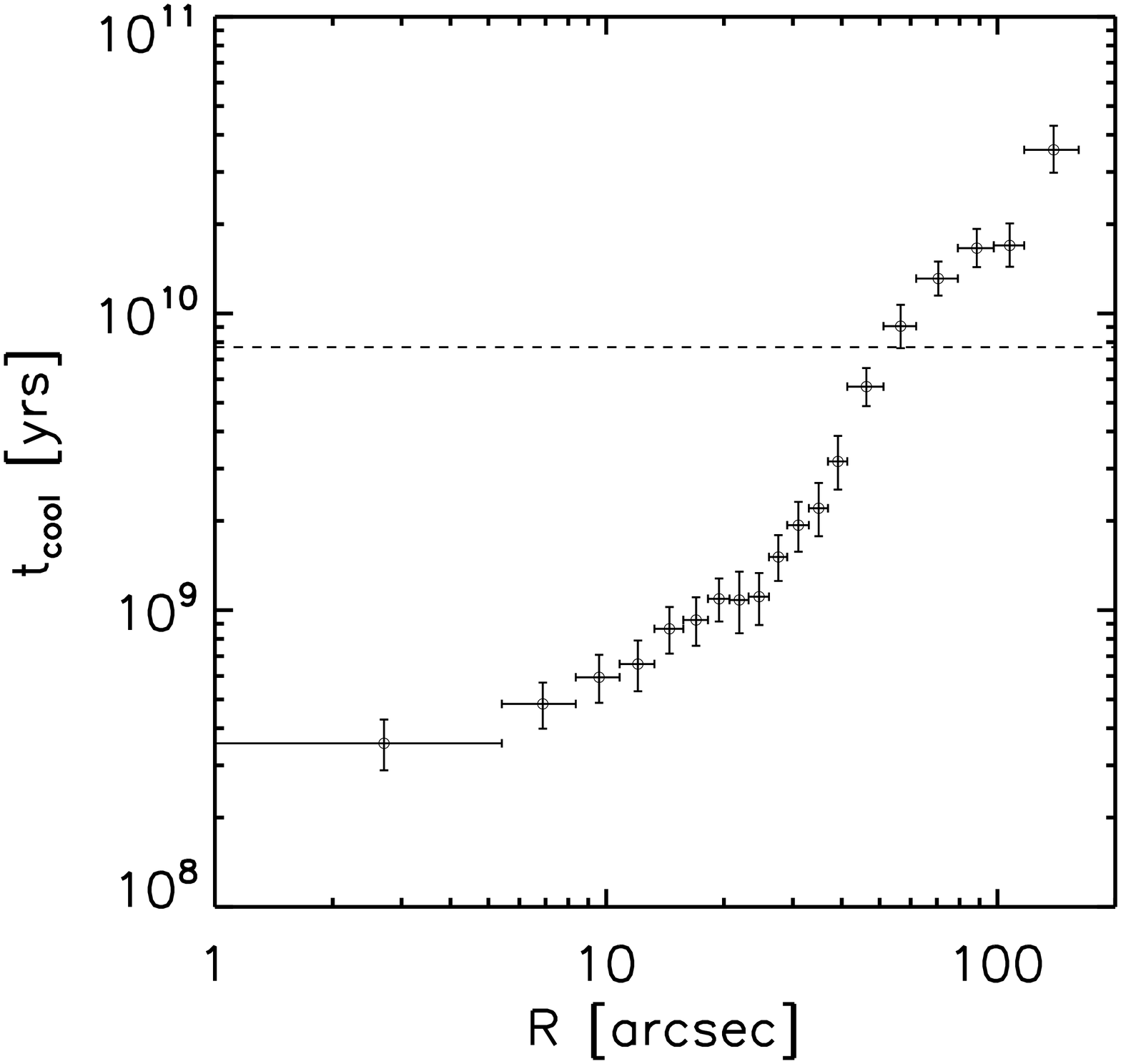} \\
\end{array}
$
\caption{{\it Upper left}: Density profile of the ICM.  {\it Upper right}: Pressure profile of the ICM.  {\it Lower left}: Entropy profile of the ICM.  {\it Lower right}: Cooling time profile of the ICM.  Bins below the dotted line (7.7$\times$10$^9$ yr) are considered to be within the cooling region.}
\end{center}
\end{figure*}

We pointed out in section 3.2.1 that the central structure is asymmetric.  A cold front could explain the structure in the global temperature profile.  We have analyzed the eastern half of the cluster, between 99 and 273 degrees from west on the sky, towards the brightness edge pointed out in Figure 2 (B).  The surface brightness profile in the direction of this region is shown in the upper left panel of Figure 10.  The break in the surface brightness profile at approximately 30$''$ corresponds to a density and temperature discontinuity shown in the upper right and lower left panel of the Figure 10.  Across the front, away from the center, we find that density falls from 0.018 cm$^{-3}$ to 0.008 cm$^{-3}$.  The temperature jumps at the front from about 3 keV to about 5.5 keV.  However, the pressure profile in the last panel of Figure 10 is continuous and shows no break.  This is characteristic of a cold front, possibly induced by sloshing of the ICM in the core due to a recent merger.  A1664 hosts a radio relic in the outer halo of the cluster \citep{gio99,gov01}.  Radio relics are thought to be a signature of mergers \citep{ens98,ens02}, strengthening the case for merger induced sloshing in A1664.

\begin{figure*}[t]
\begin{center}
$\begin{array}{cc}
\leavevmode \epsfysize=8cm \epsfbox {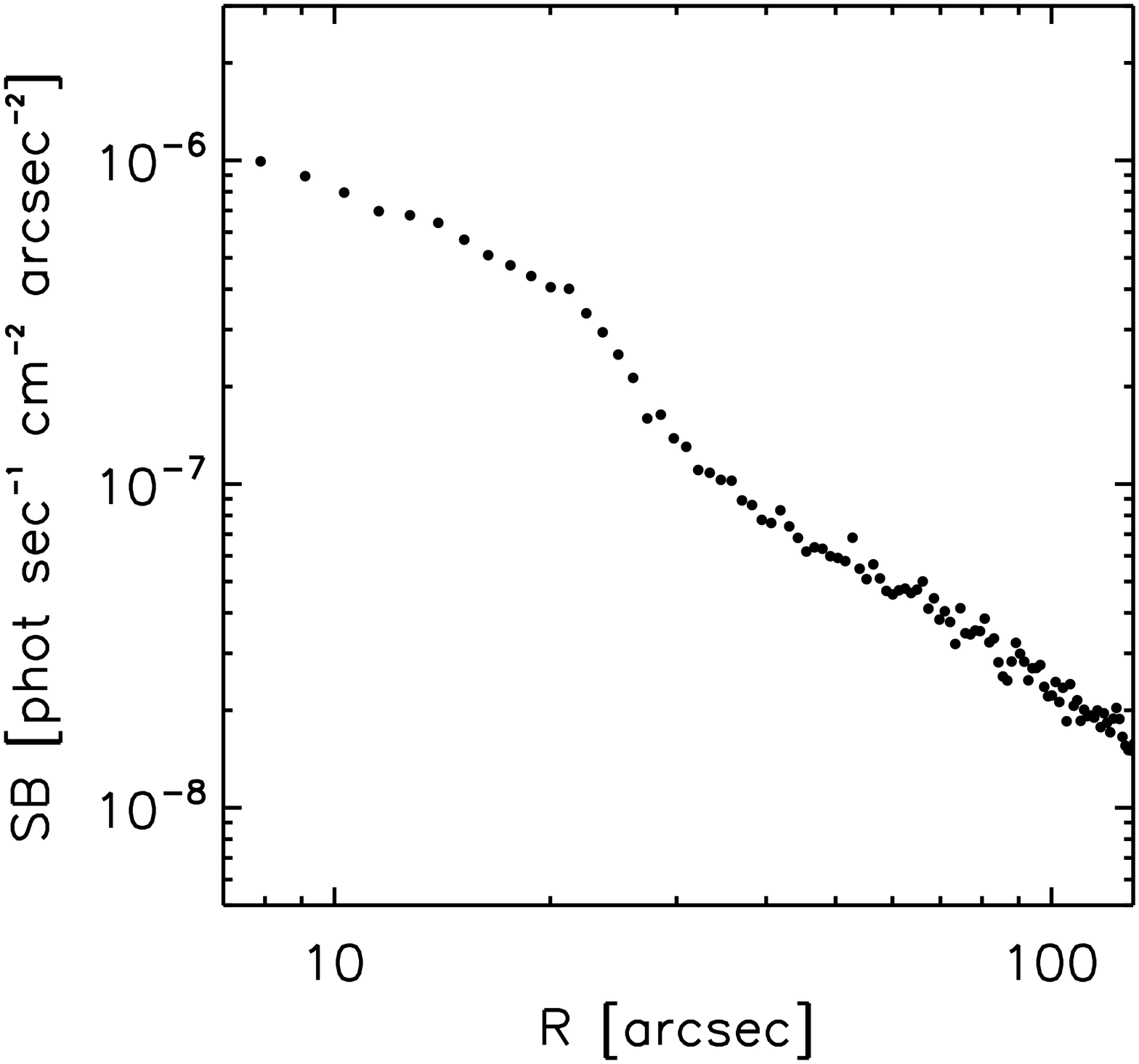} &
\leavevmode \epsfysize=8cm \epsfbox {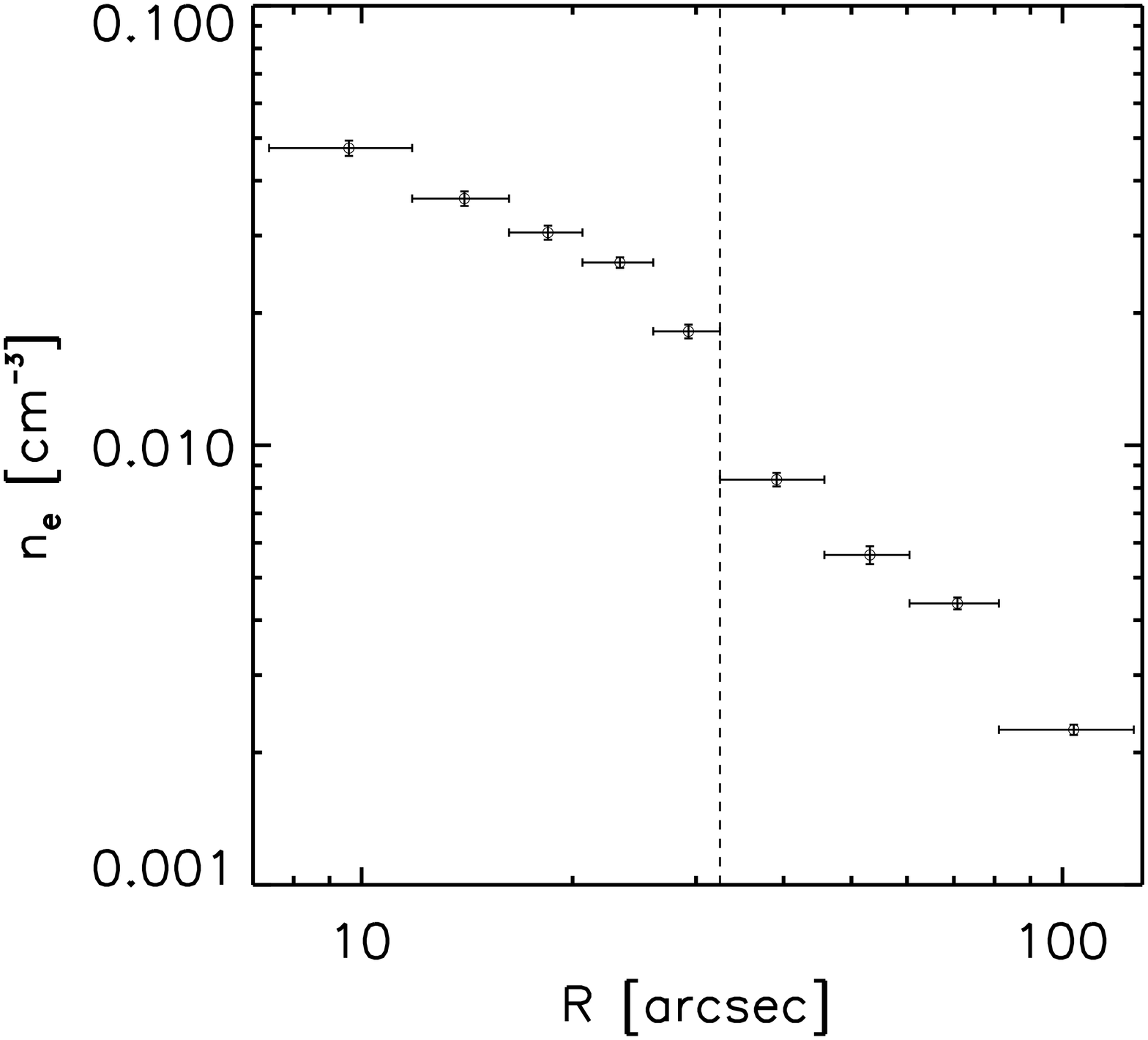} \\
\leavevmode \epsfysize=8cm \epsfbox {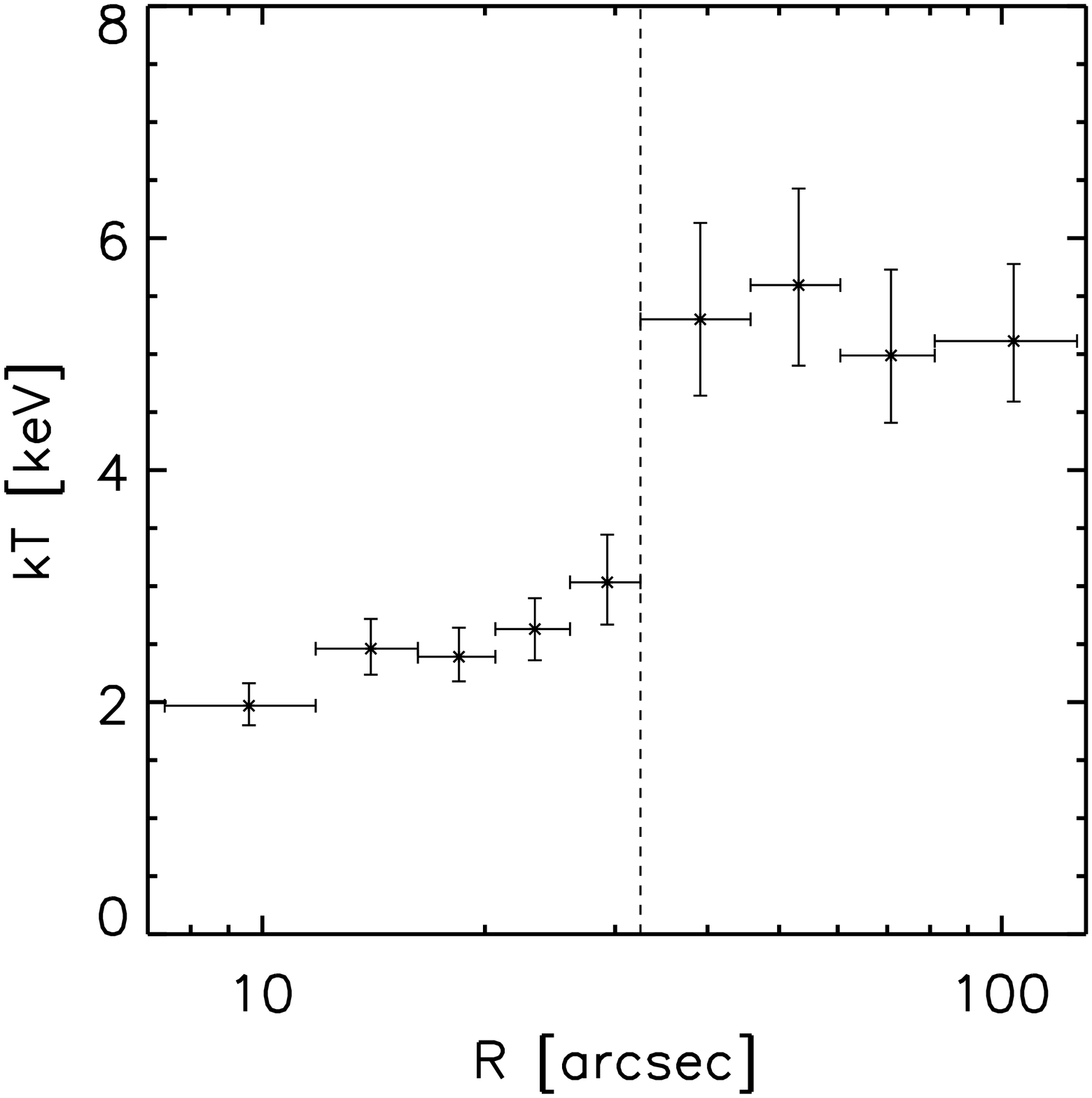} & 
\leavevmode \epsfysize=8cm \epsfbox {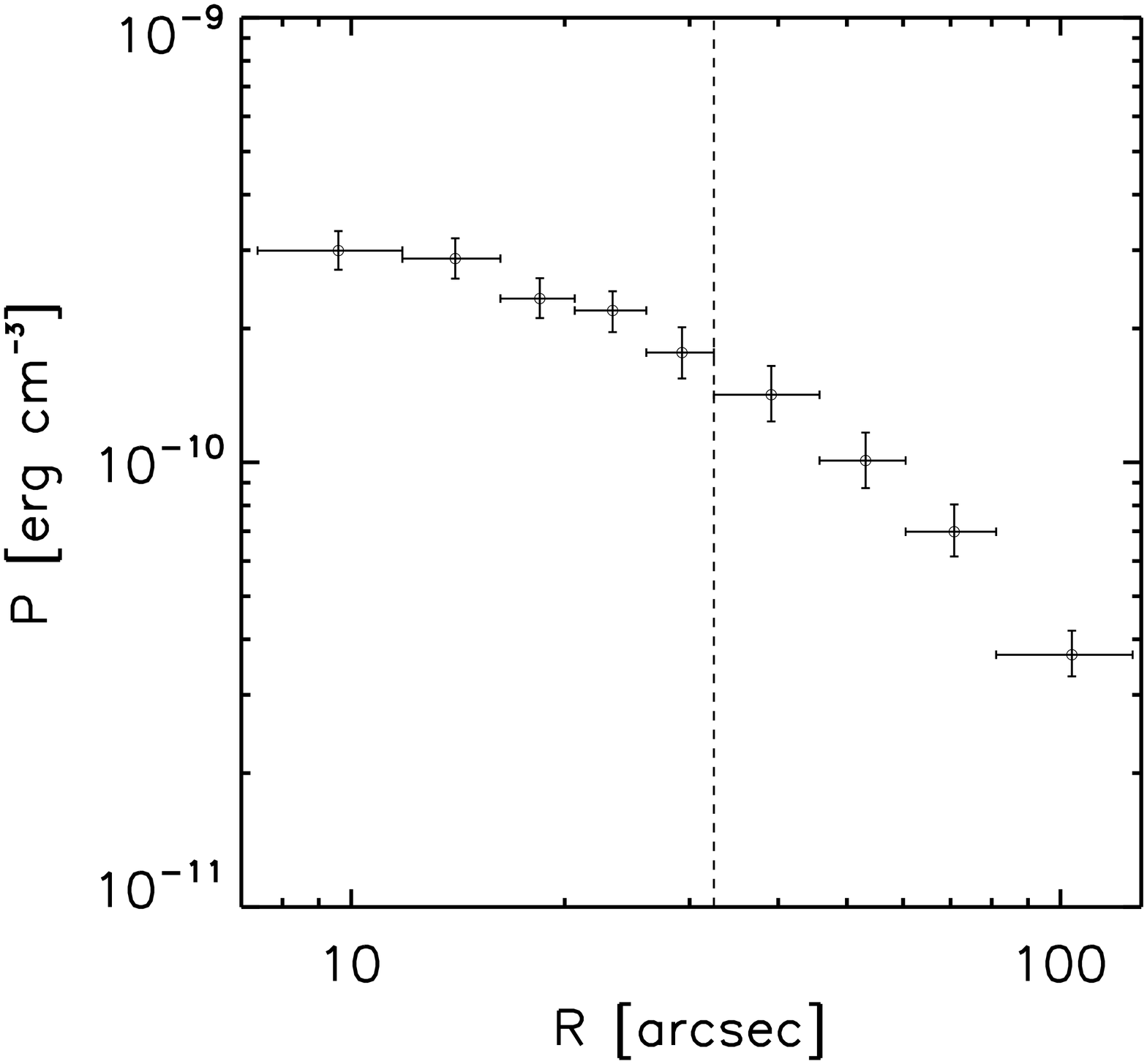} \\
\end{array}
$
\caption{{\it Upper left}: X-ray surface brightness profile of the eastern half of A1664 in the 0.4 - 7.0 keV band.  {\it Upper right}: Eastern half density profile of the ICM.  The vertical line at $\sim$ 30$''$ represents where the surface brightness break is located.  The density is discontinuous at this point as well.  {\it Lower left}: Eastern half temperature profile of the ICM.  The temperature sees a $\sim$ 2 keV jump at the break.  {\it Lower right}: Eastern half pressure profile of the ICM.  The pressure remains continuous across the surface brightness break.}
\end{center}
\end{figure*}

\subsection{Star Formation in the BCG}

The visual light of A1664's BCG, shown in Figure 2, is centered at R.A. =  $13^h03^m42^s.54$, DEC = $-24^{\degr}14'42''.05$, which lies 2.18'' NE from the X-ray centroid.  The $R$- and $U$-band surface brightness profiles are presented in the left panel of Figure 11.  We constructed these profiles using the {\it ellipse} routine in IRAF with elliptical annuli of ellipticity 0.46 and position angle 10$^{\degr}$ from north on the sky, based on the best fit to the BCG envelope.  Nearby galaxies projected onto the envelope were masked out to avoid light contamination.  The images were flux calibrated using Landolt standards.  The Galactic foreground extinction correction was calculated following \citet{car89}, assuming a foreground color excess of E(\bv) = 0.141.  Evolution and K-corrections from \citet{pog97} were applied in both filters as -0.143 and 0.130 magnitude in $R$ and -0.161 and 0.328 magnitude in $U$, respectively.  The error bars represent the statistical errors and the dashed lines represent the systematic error confidence intervals.  The method of determining these is described in \citet{bmc92}.

In the right panel of Figure 11, we present the \ur~color profile.  The inner 10$''$ shows a positive gradient (bluer toward the center) with a central rest-frame color of 1.1.  The envelope of the BCG reddens to a color of approximately 2.3.  The central color is about one magnitude bluer than a normal giant elliptical, which have colors of 2.3 - 2.6 \citep{pel90}.  This indicates star formation.

The $U$-band image, after subtraction by a smooth stellar model, is presented in the left panel of Figure 12.  Two distinct peaks can be seen in this image north and south of the center.  To the right of this in Figure 12, we show the color map on the same scale.  The normal colors are in white and the unusually blue regions are blue.  The blue features are concentrated to the center with trails of emission leading north, south, and east.  The trails to the north and south appear to end at or near neighboring galaxies.  The two galaxies labeled 1 and 2 in Figure 2 are both red and have an average color of 2.5 and 2.2 respectively.  Galaxy 1 has an absolute visual magnitude of approximately -22.7 (2L$_\star$) and galaxy 2 is approximately -21.1 (0.5L$_\star$).  These appear to be giant ellipticals which are normally devoid of cold gas, so their relationship to the star formation in the BCG, if any, is ambiguous.  They are unlikely donors of the $\sim$ 10$^{10}$ M$_\sun$ of cold gas feeding star formation in the BCG.

\begin{figure*}
\begin{center}
$\begin{array}{cc}
\leavevmode \epsfysize=8cm \epsfbox {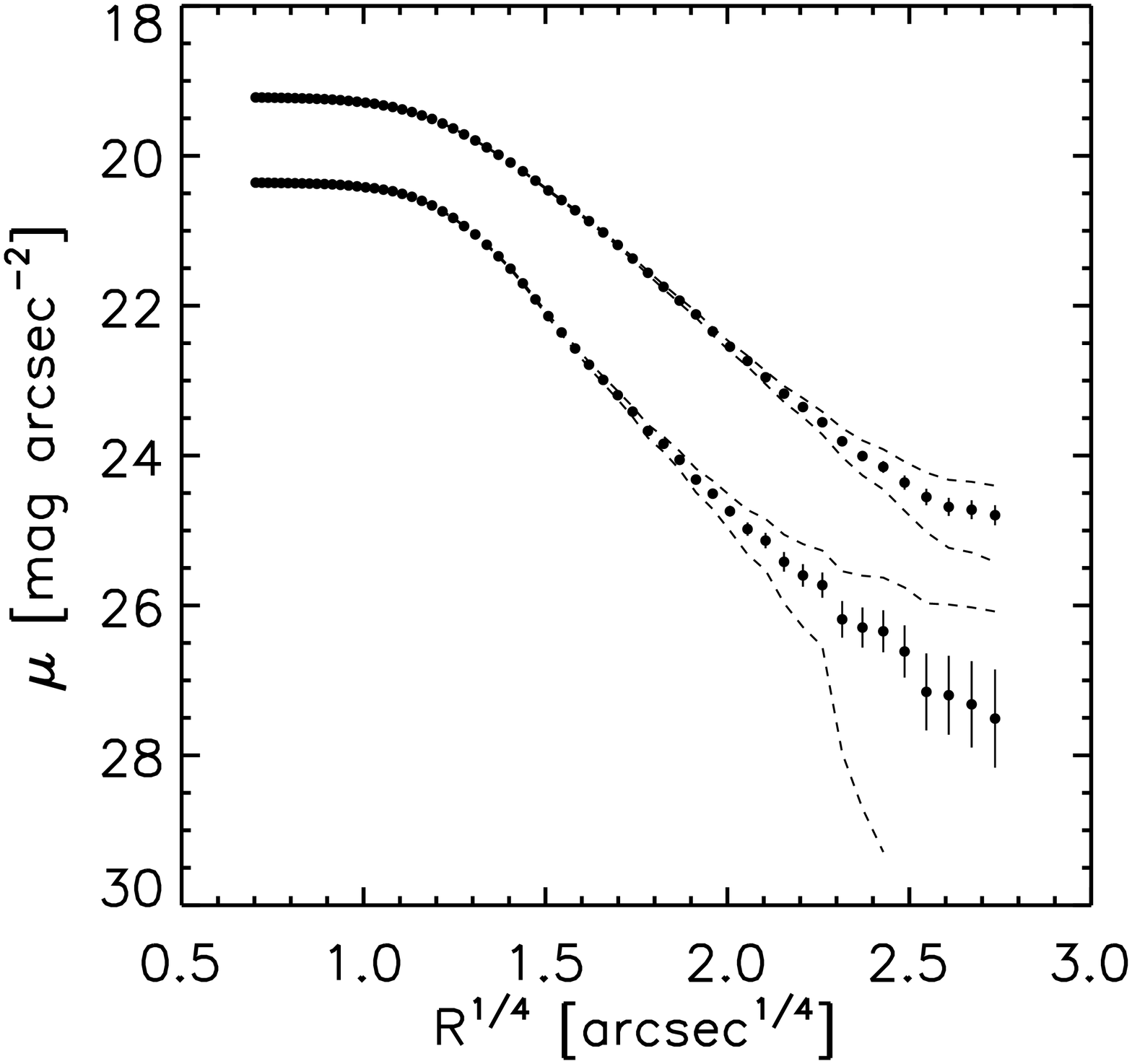} &
\leavevmode \epsfysize=8cm \epsfbox {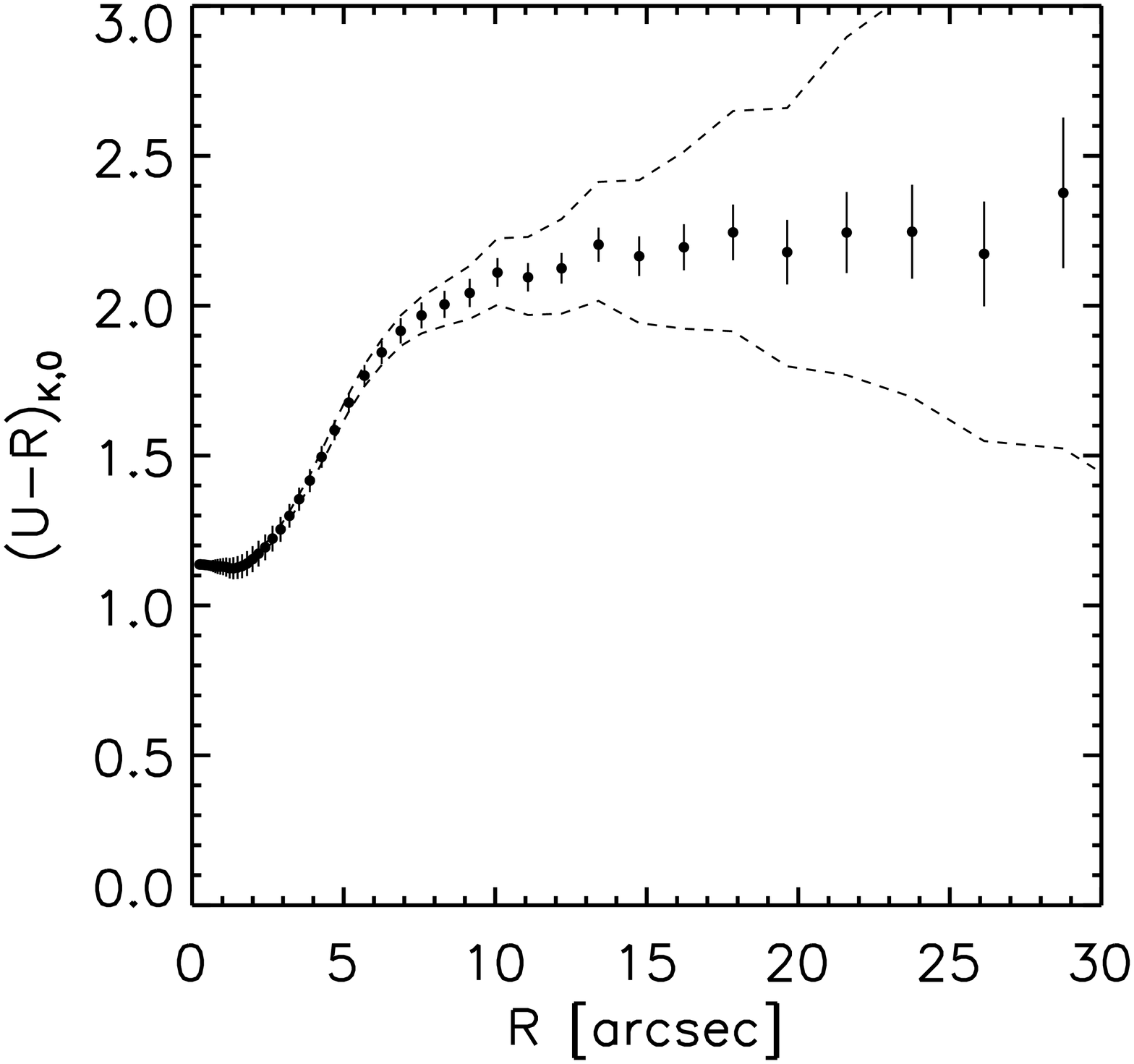} \\
\end{array}
$
\caption{{\it Left}: $R$- (upper) and $U$-band (lower) surface brightness profiles of the BCG.  Statistical errors are imperceptible at most points.  The dashed lines represent the systematic uncertainty associated with the sky background subtraction.  {\it Right}: \ur~color profile of the BCG illustrating the central $\sim$ 10$''$ starburst region.  Statistical errors and systematic error envelope are shown.}
\end{center}
\end{figure*}

\begin{figure*}
\begin{center}
$\begin{array}{cc}
\leavevmode \epsfysize=8cm \epsfbox {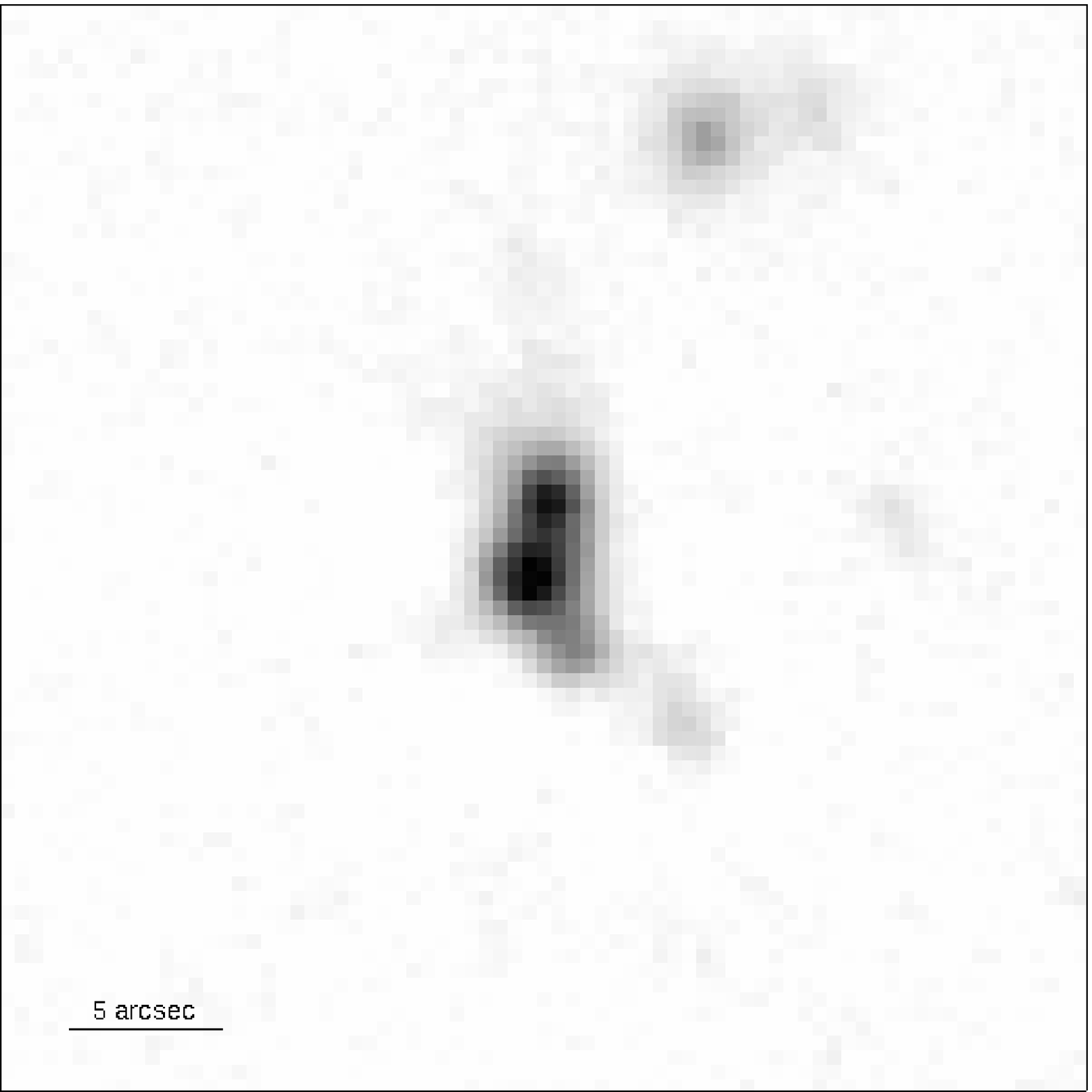} &
\leavevmode \epsfysize=8cm \epsfbox {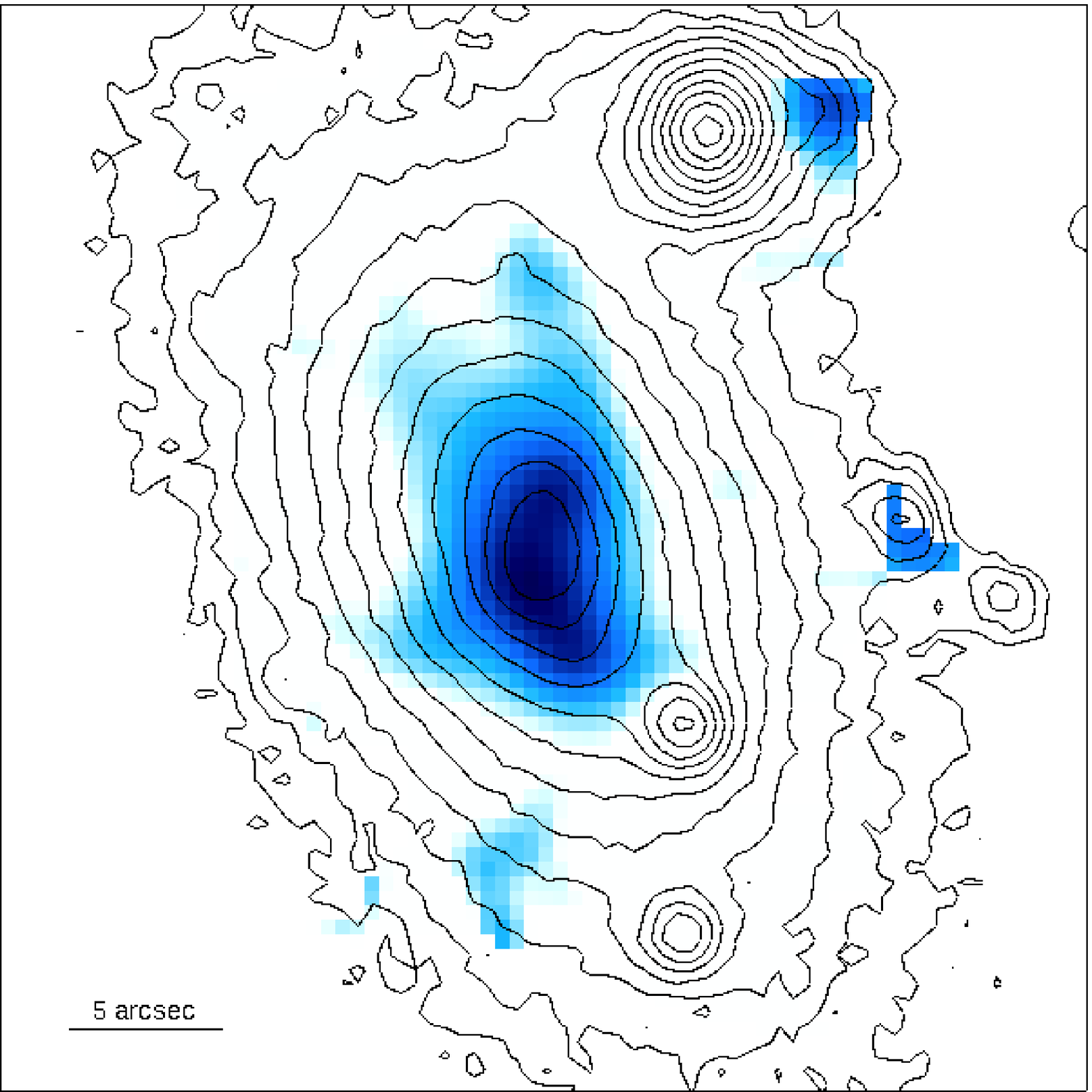} \\
\end{array}
$
\caption{{\it Left}: $U$-band image after subtracting the smooth model galaxy.  This image illustrates the separate star forming clumps.  {\it Right}: \ur~color map with $R$-band contours overlaid.  The blue regions are shown in blue; the normal colored regions are in white.  Four blue trails can be seen leading away from the center.  North is towards the top; east is towards the left.}
\end{center}
\end{figure*}

\section{Discussion}

\subsection{AGN Feedback}

AGN periodically inject energy to heat the ICM and quench cooling of the surrounding medium \citep{bir04,dun06,raf06}.  To determine whether heating is dominating in A1664, we compare the mechanical power of the AGN to the cooling power of the ICM.  AGN power can be evaluated as the $pV$ work required to inflate a cavity against the surrounding pressure of the hot ICM.  In A1664, there are no detected cavities in the X-ray.  We can, however, estimate the AGN power using the scaling relations found by \citet{bir08} between jet mechanical power and radio synchrotron power.  These relations were derived using a sample of radio galaxies with X-ray cavities.  All results referred to in this discussion are presented in Table 5.

The radio flux emerging from the nucleus of the BCG was obtained from the NVSS catalogue.  The source is unresolved in the 45$''$ FWHM beam.  Its 1.4 GHz flux 36.4 mJy corresponds to a power of 1.57$\times$10$^{24}$ W Hz$^{-1}$.  The bulk of the energy from FR I radio sources emerges in the form of mechanical energy.  Using the 1.4 GHz relationship of \citet{bir08},

\begin{equation}
\log P_{{\rm cav}} = (0.35\pm0.07) \log P_{1400} + (1.85\pm0.10),
\end{equation}
where P$_{{\rm cav}}$ has units of 10$^{42}$ erg s$^{-1}$ and P$_{1400}$ has units of 10$^{24}$ W Hz$^{-1}$, we estimate the jet power to be 8.29$\times$10$^{43}$ erg s$^{-1}$.

\begin{deluxetable*}{ccccccccc}
\tablecolumns{9}
\tablewidth{0pc}
\tablecaption{Derived Properties from X-ray \& Radio Data}
\tablehead{
	\colhead{S$_{1400}$} & \colhead{P$_{1400}$} & \colhead{S$_{352}$} & \colhead{P$_{352}$} & \colhead{P$_{\rm cav}$\tablenotemark{a}} & \colhead{P$_{\rm cav}$\tablenotemark{b}} & \colhead{L$_{X}$} & \colhead{L$_{\rm CF}$} & \colhead{r$_{\rm cool}$} \\
	\colhead{(mJy)} & \colhead{(10$^{24}$ W Hz$^{-1}$)} & \colhead{(mJy)} & \colhead{(10$^{24}$ W Hz$^{-1}$)} & \colhead{(10$^{42}$ erg s$^{-1}$)} & \colhead{(10$^{42}$ erg s$^{-1}$)} & \colhead{(10$^{42}$ erg s$^{-1}$)} & \colhead{(10$^{42}$ erg s$^{-1}$)} & \colhead{(kpc)}
 	}
\startdata
	36.4$\pm$1.2 & 1.57$\pm$0.05 & 329$\pm$9 & 14.2$\pm$0.4 & 82.9$\pm$19.4 & 66.7$\pm$29.8 & 291$\pm$6 & 46.5$\pm$10.8 & 117 \\
\enddata
\tablenotetext{a}{$\nu=1400$ MHz}
\tablenotetext{b}{$\nu=352$ MHz}
\end{deluxetable*}

\citet{raf08} have shown that star formation is predominately found in systems where the X-ray cooling luminosity is greater than the jet power.  To test this in A1664, we derived the X-ray luminosity by fitting the cooling region with a single temperature MEKAL model and using the bolometric flux (0.001-100.0 keV) of the model to calculate the luminosity.  Similarly, the cooling flow luminosity was calculated from the flux of an additional MKCFLOW component with the lower temperature set to 0.1 keV.  The cooling luminosity, 2.45$\pm$0.12$\times$10$^{44}$ erg s$^{-1}$, is calculated as the the difference between these two values.  The cooling luminosity here is three times greater than the estimated jet mechanical power found from B{\^i}rzan's relation.

The uncertainty in the jet power is large.  We can check our conclusions by using radio power measurements at a lower frequency.  The radio flux from the nucleus of the BCG in the 352 MHz WISH catalogue is 329 mJy {\citep{deb02}.  With this we can independently check the the cavity power estimated using the 1.4 GHz flux.  Using the 327 MHz relationship of \citet{bir08} (where we assume the difference in frequency dependent flux is negligible),
\begin{equation}
\log P_{{\rm cav}} = (0.62\pm0.08) \log P_{327} + (1.11\pm0.17),
\end{equation}
we find a cavity power of 6.67$\times$10$^{43}$ erg s$^{-1}$.  This value is consistent with the cavity power found with the 1.4 GHz flux.  An additional constraint is found using B{\^i}rzan's relationship between jet power and lobe radio luminosity corrected for break frequency to account for aging effects,
\begin{multline}
\log P_{{\rm cav}} = (0.53\pm0.09) \log L_{{\rm rad}} -\\ (0.74\pm0.26) \log \nu_C + (2.12\pm0.19).
\end{multline}
L$_{{\rm rad}}$ and P$_{{\rm cav}}$ are in units of 10$^{42}$ erg s$^{-1}$ and $\nu_C$ is in units of GHz.  Correcting for the break frequency reduces the scatter in this relationship by $\sim$ 50\% compared to the monochromatic relations.  The VLSS survey gives a 74 MHz flux for A1664 of 1.16 Jy.  Combining this flux measurement with the previous ones, we calculated the total radio luminosity by integrating the flux from 10 MHz to 10 GHz as two broken power law spectra, with spectral indices of 0.81 between 10 MHz and 352 MHz and 1.59 between 352 MHz and 10 GHz.  The total radio luminosity is found to be 2.01$\pm$0.05$\times$10$^{41}$ erg s$^{-1}$.  We are using the total flux (lobe plus some core flux), so this should be considered an upper limit to the lobe luminosity.  The shape of the spectrum is defined by just three flux measurements.  We have therefore taken the intermediate flux at 352 MHz to be the rough estimate of the break frequency.  Based on \citet{bir08}, this is a reasonable lower limit.  Adopting these assumptions, we estimate the upper limit on P$_{{\rm cav}}$ to be 1.80$\times$10$^{44}$ erg s$^{-1}$.  Again, this result is consistent with the cooling luminosity being greater than the energy output by the AGN.

The scatter in the B{\^i}rzan power relations is larger than the calculated difference between the jet power and the cooling luminosity.  We cannot exclude the possibility that the AGN power exceeds the cooling luminosity.  However, taking these figures at face value, the B{\^i}rzan relations give a jet power that is consistently below the cooling luminosity.  With this, the absence of detected cavities, and evidence for cooled gas in gravitational free-fall at the center \citep{wil09}, it appears that the AGN is currently underpowered compared to the cooling luminosity.  The hot atmosphere is now in a cooling phase.

\subsection{Criteria For Cooling Flow Driven Star Formation}

Since A1664 has an unusually blue core in the BCG with a star formation rate of 23 M$_\sun$ yr$^{-1}$, we have investigated whether gas cooling out of the ICM is directly feeding the cold gas reservoir and fueling star formation.  We tested A1664 against the criteria of \citet{raf08} for star formation that require the jet power to be less than the X-ray cooling luminosity, a central cooling time less than 5$\times$10$^{8}$ yr or a minimum entropy threshold of 30 keV cm$^2$, and a maximum of 20 kpc separation between the optical and X-ray centroids.  We have shown in previous sections that A1664 has an underpowered jet as well as a central cooling time and entropy of 3.5$\times$10$^{8}$ yr and 10.4 keV cm$^2$, respectively.  The X-ray and optical centroids calculated have a separation of $\sim$ 5 kpc, which satisfies all of the criteria.  

There are two other indications we have explored that star formation is related to the cooling flow (X-ray cusp).  Clusters below the entropy threshold are more likely to have high luminosity H$\alpha$ emission \citep{cav08,voit08}, a good tracer of star formation.  A1664 is one the brightest H$\alpha$ emitters in the ROSAT brightest cluster sample \citep{cra99}.  Also, the derived upper limit on the mass condensation rate (56 M$_\sun$ yr$^{-1}$) is larger than the star formation rate, although they are consistent to within their uncertainties.  \citet{raf06} found that the average ratio of mass condensation to star formation was approximately four.  For A1664, the ratio is smaller ($\sim$ 2.4).  The upper limit on cooling being greater than the star formation rate is consistent with a cooling flow driven starburst.
 
\subsection{Gas Deposited by Stripping}

Cold gas and star formation in BCGs are found almost exclusively in clusters with cooling flows (i.e., central cusp of dense gas).  Therefore, the origin of this cold gas must be linked in some way to the presence of the X-ray cusp.  Throughout this paper we have argued that this link is through condensation from the dense cool gas.  Another possible link could be through ram pressure stripping.  Because of the high density cusp, the cross section to ram pressure stripping is larger in the center than in the cluster outskirts.  Stripping could then be strong enough to drive dense molecular gas out of a gas-rich galaxy plunging through the cusp and depositing it onto the BCG.

Blue streams seen in Figure 12 could plausibly be star formation in cold gas stripped from plunging galaxies.  Two of the blue trails appear to end near red ellipticals, so their origin is unclear.  For the case of the southern most trail, \citet{wil06} has suggested that it has been dynamically induced by a disturbance from the nearby galaxy (point 2 in Figure 2).

\begin{figure}[b]
\epsscale{1.25}
\plotone{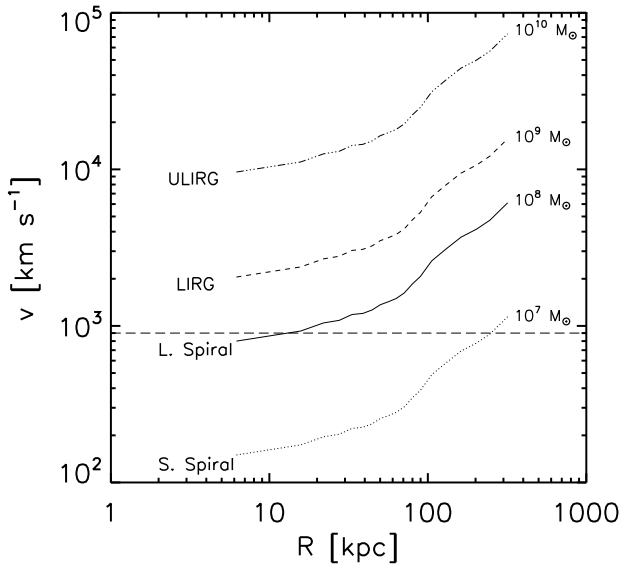}
\caption{The velocity required at varying radii in A1664 for the ram pressure stripping force to overcome the maximum gravitational restoring force of dense molecular gas in different galaxies.  From top to bottom, the lines represent typical ULIRG, LIRG, large spirals, and small spirals.  Along the right side of the plot are the average masses of molecular gas found in the inner dense region of the galaxies.  Across this range, galaxies towards the bottom are easier to strip, but have less gas to contribute to the BCG.  Galaxies towards the top are harder to strip and are rare.  The dashed horizontal line is the velocity dispersion of A1664.}
\end{figure}

We will now consider the possibility that the 10$^{10}$ $M_\sun$ cold gas reservoir was built up from gas stripped from spiral or gas-rich early-type galaxies.  Previous observations have shown that lower density cold gas from a galaxy begins to be stripped at about 0.6 to 1 Mpc from the center of a cluster \citep{ken99,vol01}.  To determine if any gas will remain in a galaxy to be deposited onto the BCG, we calculated whether a face-on, high density disk of molecular gas can survive a close passage to the center of the cluster.  In order to find the radius when stripping should occur, we compare the ram pressure stripping force to the maximum gravitational restoring force of the galaxy parallel to the flow as:
\begin{equation}
\rho v^2 = \frac{\partial \Phi(\varpi,z)}{\partial z} \sigma_{\rm s}.
\end{equation}
Here, $\varpi$ is the radius of the cold gas disk, $z$ is the distance perpendicular to the disc, $\rho$ is the density of the ICM, $v$ is the velocity of the galaxy through the ICM, and $\sigma_{\rm s}$ is the surface density of the cold gas disk.  Assuming the the galaxy has a NFW potential \citep{nav97}, the specific form for our model galaxy is,
\begin{equation}
\Phi(r) = -\frac{GM\ln(1+\frac{r}{r_{\rm s}})}{r(\ln(1+c)-\frac{c}{1+c})},
\end{equation}
where $r=\sqrt{\varpi^2+z^2}$, $M$ is the mass enclosed at $r_{200}$, $r_{\rm s}$ is the scale radius, and $c$ is the ratio of $r_{200}$ to the scale radius.  For our calculations we assume this ratio to be 10 to 1.  Solving equation 5 for the velocity a galaxy must travel to overcome the maximum gravitational restoring force per unit area,
\begin{equation}
v^2 = \frac{G M z \sigma_{\rm s}}{r^2 \rho (\ln(1+c)-\frac{c}{1+c})} \left( \frac{\ln(1+\frac{r}{r_{\rm s}})}{r}-\frac{1}{r+r_{\rm s}} \right),
\end{equation}
we present in Figure 13 how this varies with cluster radius for different galaxy types, assuming the density profile of A1664 (Figure 9).

The dotted line in Figure 13 represents a typical small spiral galaxy ($M=10^{11}$ $M_\sun$, $\sigma_{\rm s} = 100$ $M_\sun$ pc$^{-2}$, $r_{200} = 200$ kpc, $\varpi = 0.3$ kpc, $z = 1.11$ kpc), such as IC 342 \citep{cros01}.  Comparing this to the cluster's velocity dispersion of 900 km s$^{-1}$ \citep{all95}, it is likely the galaxy would lose its gas before reaching the center due to the low threshold value, making it nearly impossible for a significant amount of gas to be deposited onto the BCG.  The solid line represents a typical large spiral galaxy ($M=8 \times 10^{11}$ $M_\sun$, $\sigma_{\rm s} = 400$ $M_\sun$ pc$^{-2}$, $r_{200} = 210$ kpc, $\varpi = 0.4$ kpc, $z = 1.37$ kpc), such as M83 \citep{cros02} or the Milky Way \citep{sco87}.  A galaxy of this type would be able to hold onto its high density gas until about 80 kpc from the center.  It is plausible then for the gas to be deposited onto the BCG.  However, according to simulations by \citet{ben05}, in-falling halos in this radial velocity range on plunging orbits are rare.  On average, a spiral galaxy contains only a few 10$^{8}$ $M_\sun$ of molecular gas in the inner 1 kpc.  There would have to be hundreds of these encounters to build up the amount of cold gas seen in A1664.

Large spiral galaxies can merge and form what is known as a luminous infrared galaxy (LIRG).  These early-type galaxies contain a potential source of cold gas that is comparable to the amount of gas found in BCGs.  The dashed line represents a typical LIRG ($M=3\times10^{12}$ $M_\sun$, $\sigma_{\rm s} = 10^3$ $M_\sun$ pc$^{-2}$, $r_{200} = 250$ kpc, $\varpi = 0.5$ kpc, $z = 1.68$ kpc), such as NGC 4194 \citep{aal00}.  For galaxies in this range, it would be nearly impossible for any of the gas to be stripped, even in the cusp of A1664.  Similarly, a typical ULIRG ($M=10^{13}$ $M_\sun$, $\sigma_{\rm s} = 10^4$ $M_\sun$ pc$^{-2}$, $r_{200} = 300$ kpc, $\varpi = 1$ kpc, $z = 2.85$ kpc), represented by the dash-dot line, would also hang onto its gas despite ram pressure stripping.  Tidal stripping would be a more viable method of removing the gas from these galaxies.  However, tidal stripping has no connection to the cooling cusp.  If galaxies are depositing 10$^{10}$ $M_\sun$ of molecular gas in this way, we should see this happening in all clusters, not just in cooling flows.

Based on these calculations, ram pressure stripping is an improbable explanation for the cold gas reservoir in A1664's BCG.  This leaves us to favor the cooling flow driven star formation scenario.

\subsection{Collapsed Cavity System}

About 70 - 75\% of cooling flow clusters have detectable X-ray cavities \citep{dun05}.  Thus, close to a third of cooling flows would either be in a state of pre- or post-AGN outburst.  When cavities are formed, they sweep up low entropy gas into dense shells that may eventually break up and sink back to the cluster center.  The X-ray bar could be the remains of collapsed cavity shells.  Assuming the mass of the bar to be the minimum mass the AGN must had displaced, and further assuming the gas density (5.4$\times$10$^{-4}$ $M_{\sun}$ pc$^{-3}$) surrounding the bar to be the original, undisturbed density of the center, the amount of energy needed to displace 3.21$\times$10$^{10}$ $M_\sun$ of gas would be 3.1$\times$10$^{59}$ erg.  This energy would correspond to two cavities 44 kpc in diameter.  Energy and sizes of this magnitude are common in clusters \citep{bmc07}.

\section{Summary}

We have shown that star formation in Abell 1664 is consistent with fueling by the cooling flow.  The BCG has a blue central color (positive color gradient), a central cooling time below 5$\times$10$^8$ yr, a central entropy less than 30 keV cm$^{-2}$, an AGN power less than the X-ray cooling luminosity, and a BCG less than 20 kpc from the X-ray core.  These properties are consistent with other star forming cooling flows \citep{raf08}.  A large 10$^{10}$ $M_{\sun}$ cold gas reservoir at the cluster center is also consistent with a cooling dominated cluster.  It is unlikely that the gas was deposited by spiral or gas-rich early-type galaxies because it would take hundreds of encounters for it to be explained by ram pressure stripping alone.  Other evidence that the cooling flow is directly feeding this reservoir comes from the measured mass condensation rate of 56 $M_\sun$ yr$^{-1}$, which compares well to the observed star formation rate of 23 $M_{\sun}$ yr$^{-1}$.

Additional evidence suggesting that A1664 is currently between AGN outbursts includes the possible collapsed cavity shell remnants we have referred to as the X-ray ``bar".  We have shown that the 3.1$\times$10$^{59}$ erg it would have taken to displace this gas would have created a pair of cavities approximately 44 kpc in diameter, typical of what is seen in galaxy clusters.

We further demonstrated that if gas in the central region is cooling at the rates estimated above, we can account for the apparent dip in the metallicity found when assuming a single temperature for the medium.

\acknowledgments

C.~C.~K. would like to thank Sean McGee, and Michael Balogh for helpful discussions.  This work was funded in part by {\it Chandra} grant G07-8122X and a generous grant from the Natural Science and Engineering Research Council of Canada.


\begin{thebibliography}{}
\bibitem[Aalto \& H{\"u}ttemeister(2000)]{aal00} Aalto, S., 
\& H{\"u}ttemeister, S.\ 2000, \aap, 362, 42
\bibitem[Abell et al.(1989)]{abe89} Abell, G.~O., Corwin, 
H.~G., Jr., \& Olowin, R.~P.\ 1989, \apjs, 70, 1
\bibitem[Allen et al.(1992)]{all92} Allen, S.~W., Edge, A.~C., Fabian, A.~C., 
B{\"o}hringer, H., Crawford, C.~S., Ebeling, H., Johnstone, R.~M., Naylor, 
T., \& Schwarz, R.~A.\ 1992, \mnras, 259, 67
\bibitem[Allen et al.(1995)]{all95} Allen, S.~W., Fabian, 
A.~C., Edge, A.~C., B{\"o}hringer, H., \& White, D.~A.\ 1995, \mnras, 275, 741
\bibitem[Ascasibar \& Markevitch(2006)]{asc06} Ascasibar, Y., \& 
Markevitch, M.\ 2006, \apj, 650, 102
\bibitem[Benson(2005)]{ben05} Benson, A.~J.\ 2005, \mnras, 
358, 551
\bibitem[Bildfell et al.(2008)]{bil08} Bildfell, C., 
Hoekstra, H., Babul, A., \& Mahdavi, A.\ 2008, \mnras, 389, 1637
\bibitem[B{\^i}rzan et al.(2004)]{bir04} B{\^i}rzan, L., 
Rafferty, D.~A., McNamara, B.~R., Wise, M.~W., 
\& Nulsen, P.~E.~J.\ 2004, \apj, 607, 800
\bibitem[B\^{i}rzan et al.(2008)]{bir08} B\^{i}rzan, L., McNamara, B.~R., 
Nulsen, P.~E.~J., Carilli, C.~L. \& Wise, M.~W. \ 2008, \apj, 686, 859
\bibitem[B{\"o}hringer et 
al.(2001)]{boh01} B{\"o}hringer, H., et al.\ 2001, \aap, 365, L181
\bibitem[Bower et al.(2006)]{bow06} Bower, R.~G., Benson, 
A.~J., Malbon, R., Helly, J.~C., Frenk, C.~S., Baugh, C.~M., Cole, S., 
\& Lacey, C.~G.\ 2006, \mnras, 370, 645
\bibitem[Br{\"u}ggen \& Kaiser(2002)]{bru02} Br{\"u}ggen, M., \& 
Kaiser, C.~R.\ 2002, \nat, 418, 301
\bibitem[Br{\"u}ggen et al.(2005)]{bru05} Br{\"u}ggen, M., 
Ruszkowski, M., \& Hallman, E.\ 2005, \apj, 630, 740
\bibitem[Buote(2000)]{buo00} Buote, D.~A.\ 2000, \mnras, 311, 
176
\bibitem[Buote(2001)]{buo01} Buote, D.~A.\ 2001, \apj, 548, 
652
\bibitem[Cardelli et al.(1989)]{car89} Cardelli, J.~A., 
Clayton, G.~C., \& Mathis, J.~S.\ 1989, \apj, 345, 245
\bibitem[Cardiel et al.(1998)]{card98} Cardiel, N., Gorgas, 
J., \& Aragon-Salamanca, A.\ 1998, \mnras, 298, 977
\bibitem[Cavagnolo et al.(2008)]{cav08} Cavagnolo, K.~W., 
Donahue, M., Voit, G.~M., \& Sun, M.\ 2008, \apjl, 683, L107
\bibitem[Churazov et al.(2002)]{chu02} Churazov, E., Sunyaev, 
R., Forman, W., \& B{\"o}hringer, H.\ 2002, \mnras, 332, 729
\bibitem[Clarke et al.(2004)]{cla04} Clarke, T.~E., Blanton, 
E.~L., \& Sarazin, C.~L.\ 2004, \apj, 616, 178
\bibitem[Crawford et al.(1995)]{cra95} Crawford, C.~S., Edge, 
A.~C., Fabian, A.~C., Allen, S.~W., Bohringer, H., Ebeling, H., McMahon, 
R.~G., \& Voges, W.\ 1995, \mnras, 274, 75
\bibitem[Crawford et al.(1999)]{cra99} Crawford, C.~S., 
Allen, S.~W., Ebeling, H., Edge, A.~C., 
\& Fabian, A.~C.\ 1999, \mnras, 306, 857
\bibitem[Crosthwaite et al.(2001)]{cros01} Crosthwaite, L.~P., 
Turner, J.~L., Hurt, R.~L., Levine, D.~A., Martin, R.~N., 
\& Ho, P.~T.~P.\ 2001, \aj, 122, 797
\bibitem[Crosthwaite et al.(2002)]{cros02} Crosthwaite, L.~P., 
Turner, J.~L., Buchholz, L., Ho, P.~T.~P., 
\& Martin, R.~N.\ 2002, \aj, 123, 1892
\bibitem[Croton et al.(2006)]{cro06} Croton, D.~J., Springel, V., White, S.~D.~M., 
De Lucia, G., Frenk, C.~S., Gao, L., Jenkins, A., Kauffmann, G., Navarro, J.~F., 
\& Yoshida, N.\ 2006, \mnras, 365, 11
\bibitem[Dalla Vecchia et al.(2004)]{dal04} Dalla Vecchia, 
C., Bower, R.~G., Theuns, T., Balogh, M.~L., Mazzotta, P., 
\& Frenk, C.~S.\ 2004, \mnras, 355, 995
\bibitem[Dennis \& Chandran(2005)]{den05} Dennis, T.~J., \& 
Chandran, B.~D.~G.\ 2005, \apj, 622, 205
\bibitem[De Breuck et al.(2002)]{deb02} De Breuck, C., Tang, 
Y., de Bruyn, A.~G., Rottgering, H., 
\& van Breugel, W.\ 2002, VizieR Online Data Catalog, 8069, 0
\bibitem[De Grandi \& Molendi(2001)]{deg01} De Grandi, S., \& 
Molendi, S.\ 2001, \apj, 551, 153
\bibitem[Dickey \& Lockman(1990)]{dic90} Dickey, J.~M., \& 
Lockman, F.~J.\ 1990, \araa, 28, 215
\bibitem[Dolag et al.(2004)]{dol04} Dolag, K., Jubelgas, M., 
Springel, V., Borgani, S., \& Rasia, E.\ 2004, \apjl, 606, L97
\bibitem[Donahue et al.(2007)]{don07} Donahue, M., Sun, M., 
O'Dea, C.~P., Voit, G.~M., \& Cavagnolo, K.~W.\ 2007, \aj, 134, 14
\bibitem[Dunn et al.(2005)]{dun05} Dunn, R.~J.~H., Fabian, 
A.~C., \& Taylor, G.~B.\ 2005, \mnras, 364, 1343
\bibitem[Dunn \& Fabian(2006)]{dun06} Dunn, R.~J.~H., 
\& Fabian, A.~C.\ 2006, \mnras, 373, 959
\bibitem[Dupke et al.(2007)]{dup07} Dupke, R., White, R.~E., 
III, \& Bregman, J.~N.\ 2007, \apj, 671, 181
\bibitem[Edge(2001)]{edge01} Edge, A.~C.\ 2001, \mnras, 328, 
762
\bibitem[Edge et al.(2002)]{edge02} Edge, A.~C., Wilman, 
R.~J., Johnstone, R.~M., Crawford, C.~S., Fabian, A.~C., 
\& Allen, S.~W.\ 2002, \mnras, 337, 49
\bibitem[Edwards et al.(2007)]{love07} Edwards, L.~O.~V., 
Hudson, M.~J., Balogh, M.~L., \& Smith, R.~J.\ 2007, \mnras, 379, 100
\bibitem[En{\ss}lin et al.(1998)]{ens98} En{\ss}lin, T.~A., Biermann, 
P.~L., Klein, U., \& Kohle, S.\ 1998, \aap, 332, 395
\bibitem[En{\ss}lin \& Br{\"u}ggen(2002)]{ens02} En{\ss}lin, T.~A., \& 
Br{\"u}ggen, M.\ 2002, \mnras, 331, 1011
\bibitem[Fabian(1994)]{fab94} Fabian, A.~C.\ 1994, \araa, 32, 
277
\bibitem[Fabian et al.(2006)]{fab06} Fabian, A.~C., Sanders, 
J.~S., Taylor, G.~B., Allen, S.~W., Crawford, C.~S., Johnstone, R.~M., \& 
Iwasawa, K.\ 2006, \mnras, 366, 417
\bibitem[Ferrarese \& Merritt(2000)]{fer00} Ferrarese, L., \& 
Merritt, D.\ 2000, \apjl, 539, L9
\bibitem[Gebhardt et al.(2000)]{geb00} Gebhardt, K., et al.\ 
2000, \apjl, 539, L13
\bibitem[Gilfanov et al.(1987)]{gil87} Gilfanov, M.~R., 
Syunyaev, R.~A., \& Churazov, E.~M.\ 1987, Soviet Astronomy Letters, 13, 3
\bibitem[Giovannini et al.(1999)]{gio99} Giovannini, G., 
Tordi, M., \& Feretti, L.\ 1999, New Astronomy, 4, 141
\bibitem[Govoni et al.(2001)]{gov01} Govoni, F., Feretti, L., 
Giovannini, G., B{\"o}hringer, H., Reiprich, T.~H., \& Murgia, M.\ 2001, 
\aap, 376, 803
\bibitem[Grevesse \& Sauval(1998)]{gre98} Grevesse, N., \& 
Sauval, A.~J.\ 1998, Space Science Reviews, 85, 161
\bibitem[H{\"a}ring \& Rix(2004)]{har04} H{\"a}ring, N., \& 
Rix, H.-W.\ 2004, \apjl, 604, L89
\bibitem[Heinz et al.(2006)]{hei06} Heinz, S., Br{\"u}ggen, 
M., Young, A., \& Levesque, E.\ 2006, \mnras, 373, L65
\bibitem[Hicks \& Mushotzky(2005)]{hic05} Hicks, A.~K., \& 
Mushotzky, R.\ 2005, \apjl, 635, L9
\bibitem[Holtzman et al.(1996)]{hol96} Holtzman, J.~A., et 
al.\ 1996, \aj, 112, 416
\bibitem[Hu et al.(1985)]{hu85} Hu, E.~M., Cowie, L.~L., 
\& Wang, Z.\ 1985, \apjs, 59, 447
\bibitem[Johnstone et al.(1987)]{john87} Johnstone, R.~M., 
Fabian, A.~C., \& Nulsen, P.~E.~J.\ 1987, \mnras, 224, 75
\bibitem[Kay et al.(2004)]{kay04} Kay, S.~T., Thomas, P.~A., 
Jenkins, A., \& Pearce, F.~R.\ 2004, \mnras, 355, 1091
\bibitem[Kenney \& Koopmann(1999)]{ken99} Kenney, J.~D.~P., \& 
Koopmann, R.~A.\ 1999, \aj, 117, 181
\bibitem[Markevitch \& Vikhlinin(2007)]{mar07} Markevitch, 
M., \& Vikhlinin, A.\ 2007, \physrep, 443, 1
\bibitem[McNamara \& O'Connell(1989)]{bmc89} McNamara, B.~R., \& 
O'Connell, R.~W.\ 1989, \aj, 98, 2018
\bibitem[McNamara \& O'Connell(1992)]{bmc92} 
McNamara, B.~R., \& O'Connell, R.~W.\ 1992, \apj, 393, 579
\bibitem[McNamara et al.(2000)]{bmc00} McNamara, B.~R., et 
al.\ 2000, \apjl, 534, L135
\bibitem[McNamara(2002)]{bmc02} McNamara, B.~R.\ 2002, The 
High Energy Universe at Sharp Focus: Chandra Science, 262, 351
\bibitem[McNamara \& Nulsen(2007)]{bmc07} McNamara, B.~R., \& 
Nulsen, P.~E.~J.\ 2007, \araa, 45, 117
\bibitem[Molendi \& Gastaldello(2001)]{mol01} Molendi, S., \& 
Gastaldello, F.\ 2001, \aap, 375, L14
\bibitem[Morris \& Fabian(2003)]{mor03} Morris, R.~G., \& 
Fabian, A.~C.\ 2003, \mnras, 338, 824
\bibitem[Navarro et al.(1997)]{nav97} Navarro, J.~F., Frenk, 
C.~S., \& White, S.~D.~M.\ 1997, \apj, 490, 493
\bibitem[O'Dea et al.(2008)]{ode08} O'Dea, C.~P., et al.\ 
2008, \apj, 681, 1035
\bibitem[Peletier et al.(1990)]{pel90} Peletier, R.~F., 
Davies, R.~L., Illingworth, G.~D., Davis, L.~E., 
\& Cawson, M.\ 1990, \aj, 100, 1091
\bibitem[Peterson et al.(2003)]{pet03} Peterson, J.~R., Kahn, 
S.~M., Paerels, F.~B.~S., Kaastra, J.~S., Tamura, T., Bleeker, J.~A.~M., 
Ferrigno, C., \& Jernigan, J.~G.\ 2003, \apj, 590, 207
\bibitem[see Peterson \& Fabian(2006)]{pet06} Peterson, J.~R., 
\& Fabian, A.~C.\ 2006, \physrep, 427, 1
\bibitem[Poggianti(1997)]{pog97} Poggianti, B.~M.\ 1997, \aaps, 122, 
399
\bibitem[Poole et al.(2008)]{poo08} Poole, G.~B., Babul, A., 
McCarthy, I.~G., Sanderson, A.~J.~R., 
\& Fardal, M.~A.\ 2008, \mnras, 391, 1163
\bibitem[Rafferty et al.(2006)]{raf06} Rafferty, D.~A., 
McNamara, B.~R., Nulsen, P.~E.~J., \& Wise, M.~W.\ 2006, \apj, 652, 216
\bibitem[Rafferty et al.(2008)]{raf08} Rafferty, D.~A., 
McNamara, B.~R., \& Nulsen, P.~E.~J.\ 2008, \apj, 687, 899
\bibitem[Reynolds et al.(2002)]{rey02} Reynolds, C.~S., 
Heinz, S., \& Begelman, M.~C.\ 2002, \mnras, 332, 271
\bibitem[Ruszkowski et al.(2004a)]{rus04a} Ruszkowski, M., 
Br{\"u}ggen, M., \& Begelman, M.~C.\ 2004, \apj, 611, 158
\bibitem[Ruszkowski et al.(2004b)]{rus04b} Ruszkowski, M., 
Br{\"u}ggen, M., \& Begelman, M.~C.\ 2004, \apj, 615, 675
\bibitem[Salom{\'e} \& Combes(2003)]{sal03} Salom{\'e}, P., \& 
Combes, F.\ 2003, \aap, 412, 657
\bibitem[Sanders \& Fabian(2002)]{san02} Sanders, J.~S., \& 
Fabian, A.~C.\ 2002, \mnras, 331, 273
\bibitem[Sanders et al.(2008)]{san08} Sanders, J.~S., Fabian, 
A.~C., Allen, S.~W., Morris, R.~G., Graham, J., 
\& Johnstone, R.~M.\ 2008, \mnras, 385, 1186 
\bibitem[Schmidt et al.(2002)]{sch02} Schmidt, R.~W., Fabian, 
A.~C., \& Sanders, J.~S.\ 2002, \mnras, 337, 71
\bibitem[Scoville \& Sanders(1987)]{sco87} Scoville, N.~Z., 
\& Sanders, D.~B.\ 1987, Interstellar Processes, 134, 21
\bibitem[Soker(2008)]{sok08} Soker, N.\ 2008, \apjl, 684, 
L5
\bibitem[Tittley \& Henriksen(2005)]{tit05} Tittley, E.~R., \& 
Henriksen, M.\ 2005, \apj, 618, 227
\bibitem[Vernaleo \& Reynolds(2006)]{ver06} Vernaleo, J.~C., \& 
Reynolds, C.~S.\ 2006, \apj, 645, 83
\bibitem[Vikhlinin(2004)]{vik04} Vikhlinin, A., "Spacial structure in the ACIS OBF contamination", 
{\it CXC internal memo}, May 9, 2004
\bibitem[Voigt \& Fabian(2004)]{voi04} Voigt, L.~M., \& 
Fabian, A.~C.\ 2004, \mnras, 347, 1130
\bibitem[Voit et al.(2002)]{voit02} Voit, G.~M., Bryan, G.~L., 
Balogh, M.~L., \& Bower, R.~G.\ 2002, \apj, 576, 601
\bibitem[Voit \& Donahue(2005)]{voit05} Voit, G.~M., \& 
Donahue, M.\ 2005, \apj, 634, 955
\bibitem[Voit et al.(2008)]{voit08} Voit, G.~M., Cavagnolo, 
K.~W., Donahue, M., Rafferty, D.~A., McNamara, B.~R., 
\& Nulsen, P.~E.~J.\ 2008, \apjl, 681, L5
\bibitem[Vollmer et al.(2001)]{vol01} Vollmer, B., Braine, J., 
Balkowski, C., Cayatte, V., \& Duschl, W.~J.\ 2001, \aap, 374, 824
\bibitem[Wilman et al.(2006)]{wil06} Wilman, R.~J., Edge, 
A.~C., \& Swinbank, A.~M.\ 2006, \mnras, 371, 93
\bibitem[Wilman et al.(2009)]{wil09} Wilman, R.~J., Edge, 
A.~C., \& Swinbank, A.~M.\ 2009, arXiv:0902.4720
\bibitem[Wise et al.(2004)]{wise04} Wise, M.~W., McNamara, 
B.~R., \& Murray, S.~S.\ 2004, \apj, 601, 184
\bibitem[Zakamska \& Narayan(2003)]{zak03} Zakamska, N.~L., \& 
Narayan, R.\ 2003, \apj, 582, 162
\end{thebibliography}
\end{document}